\documentclass[twocolumn]{aastex63}
\pdfoutput=1

\newcommand\Gaia{{\it Gaia}}
\newcommand\tess{{\it TESS}}
\newcommand\teff{{\log T_{\rm {eff}}}}
\newcommand\lum{{\log L/L_\odot}}

\usepackage{amsmath,graphicx,multirow}
 
\received{}
\revised{}
\accepted{}
\submitjournal{ApJ}

\shorttitle{Pulsations in Evolved Supergiants}
\shortauthors{Dorn-Wallenstein et al.}

\begin{document}

\title{The Properties of Fast Yellow Pulsating Supergiants: \\ FYPS Point the Way to Missing Red Supergiants}

\correspondingauthor{Trevor Z. Dorn-Wallenstein}
\email{tdorn-wallenstein@carnegiescience.edu}

\author[0000-0003-3601-3180]{Trevor Z. Dorn-Wallenstein}\thanks{Carnegie Fellow}
\affiliation{University of Washington Astronomy Department \\
Physics and Astronomy Building, 3910 15th Ave NE  \\
Seattle, WA 98105, USA} 
\affiliation{Observatories of the Carnegie Institution for Science \\ 813 Santa Barbara Street \\ Pasadena, CA 91101, USA}

\author[0000-0003-2184-1581]{Emily M. Levesque}
\affiliation{University of Washington Astronomy Department \\
Physics and Astronomy Building, 3910 15th Ave NE  \\
Seattle, WA 98105, USA}


\author[0000-0002-0637-835X]{James R. A. Davenport}
\affiliation{University of Washington Astronomy Department \\
Physics and Astronomy Building, 3910 15th Ave NE  \\
Seattle, WA 98105, USA} 

\author[0000-0002-5787-138X]{Kathryn F. Neugent}
\affiliation{Dunlap Institute for Astronomy \& Astrophysics, University of Toronto \\ 50 St. George Street \\ Toronto, ON M5S 3H4, Canada}
\affiliation{Lowell Observatory, 1400 W Mars Hill Road, Flagstaff, AZ 86001}

\author[0000-0003-2528-3409]{Brett M.~Morris}
\affiliation{Center for Space and Habitability, University of Bern, \\ Gesellschaftsstrasse 6, 3012 Bern, Switzerland}

\author[0000-0002-4924-444X]{K. Azalee Bostroem}\thanks{LSSTC Catalyst Fellow}
\affiliation{Steward Observatory, University of Arizona \\
933 North Cherry Avenue  \\
Tucson, AZ 85721-0065, USA}


\begin{abstract}

Fast yellow pulsating supergiants (FYPS) are a recently-discovered class of evolved massive pulsator. As candidate post-red supergiant objects, and one of the few classes of pulsating evolved massive stars, these objects have incredible potential to change our understanding of the structure and evolution of massive stars. Here we examine the lightcurves of a sample of 126 cool supergiants in the Magellanic Clouds observed by the Transiting Exoplanet Survey Satellite (\tess~) in order to identify pulsating stars. After making quality cuts and filtering out contaminant objects, we examine the distribution of pulsating stars in the Hertzprung-Russel (HR) diagram, and find that FYPS occupy a region above $\lum \gtrsim 5.0$. This luminosity boundary corresponds to stars with initial masses of $\sim$18-20 $M_\odot$, consistent with the most massive red supergiant progenitors of supernovae (SNe) II-P, as well as the observed properties of SNe IIb progenitors. This threshold is in agreement with the picture that FYPS are post-RSG stars. Finally, we characterize the behavior of FYPS pulsations as a function of their location in the HR diagram. We find low frequency pulsations at higher effective temperatures, higher frequency pulsations at lower temperatures, with a transition between the two behaviors at intermediate temperatures. The observed properties of FYPS make them fascinating objects for future theoretical study.

\end{abstract}

\section{Introduction} \label{sec:intro}

In the standard evolutionary scenario \citep{conti75}, massive stars with initial masses between 8 and 25 $M_\odot$ end their lives as red supergiants (RSG), which explode as type II-P or II-L supernovae (SNe). In this picture, more massive stars partially or completely lose their envelopes through stellar winds. They then end their lives as yellow or blue supergiants in ``stripped envelope'' SNe (SESNe) of types IIb, Ib, and Ic \citep[e.g.,][]{filippenko97}, with the progression through these subtypes indicating the degree of envelope stripping. 

While RSGs with initial masses up to $\sim$25 $M_\odot$ are observed around the local group \citep[e.g.][]{levesque05}, actual observations of SNe II-P progenitors with pre-explosion imaging have revealed a dearth of high-mass RSG progenitors of type II SNe \citep[][]{smartt09,smartt15}. This {\it red supergiant problem} has continued to worsen since its discovery: the latest Bayesian analyses conclude that SNe II-P can only be produced by massive stars with initial masses beneath a limit $M_h$ that is somewhere in the vicinity of 20 $M_\odot$ at the highest \citep[][]{kochanek20}, and the low X-ray luminosities of SNe II-P are inconsistent with the high mass loss rates observed in high mass RSGs \citep{dwarkadas14}. One proposed solution to the RSG problem is that the RSGs with initial masses $M_i\ge M_h$ lose enough of their envelopes through mass loss or interactions with a binary companion to evolve blueward in the HR diagram before exploding \citep[e.g.][]{ekstrom12,neugent20}, possibly as SESNe. Indeed, such enhanced mass loss rates are observed in massive RSGs \citep{humphreys20}, and may also be necessary to explain the observed diversity of SNe II \citep{martinez22}. 

Validating this scenario requires compiling a sample of post-red supergiant objects (post-RSGs) large enough to determine the minimum luminosity (and thus minimum initial mass) star that produces a post-RSG. There is a rich history of work identifying the most luminous post-RSGs via their circumstellar material \citep[e.g.][]{jones93,humphreys97,humphreys02,shenoy16}, photometric variability induced by dynamical instabilities \citep[e.g.][]{nieuwenhuijzen95,stothers01} or spectroscopic signatures \citep[e.g.][]{humphreys13,gordon16,kourniotis22}. However, less-luminous post-RSGs are quite difficult to distinguish from their pre-RSG counterparts with otherwise-identical temperatures and luminosities. 

The {\it Transiting Exoplanet Survey Satellite} (\tess, \citealt{ricker15}) has opened a new window into understanding massive stars. By observing at two-minute cadence from space, \tess~has revealed multiple modes of high-frequency variability in evolved massive stars that cannot be detected from ground-based surveys. An exciting new result has been the discovery of a new class of evolved massive pulsator: fast yellow pulsating supergiants (FYPS; \citealt{dornwallenstein20b}). Massive stars are not expected to pulsate as they first cross the HR diagram; however, post-RSGs may pulsate after a sufficient amount of their envelope has been lost \citep[e.g.][]{saio13}. With a large enough sample of FYPS, the evolutionary status of this new class of pulsators can be confirmed, and their properties studied. Finally, using FYPS pulsations to probe the interior of an evolved massive star via asteroseismology would be a unique and incredibly powerful constraint on the late stages of stellar evolution, and we wish to infer the feasibility of performing such an analysis.

In this paper we search for pulsations in the \tess~lightcurves of a sample of 201 cool supergiants in the Magellanic Clouds. We describe our sample selection procedure as well as how we identify new FYPS in \S\ref{sec:data}. We explore the distribution of pulsators in the upper HR diagram in \S\ref{sec:results}, and discuss the evolutionary status of FYPS as well as their suitability for asteroseismic studies in \S\ref{sec:discussion}. When then summarize our key results and conclude in \S\ref{sec:conclusion}.

\section{Methodology}\label{sec:data}

\subsection{Sample Selection}

As in \citet{dornwallenstein20b}, we use the sample of yellow supergiants (YSGs) from \citet{neugent10,neugent12_ysg}, who used spectra obtained with the Hydra multi-object spectrograph on the Cerro Tololo 4-meter telescope to confirm the membership of a large sample of YSGs and RSGs in the Large Magellanic Cloud (LMC) and Small Magellanic Cloud (SMC), along with updated formulae derived from Kurucz \citep{kurucz92} and MARCS \citep{gustafsson08} to obtain effective temperatures ($\teff$) and luminosities ($\lum$) from near-infrared photometry. These measurements have a typical precision of 0.015 dex and 0.10 dex in $\teff$ and $\lum$ respectively. While other, more complete catalogs of Magellanic Cloud supergiants exist (e.g., \citealt{yang19,yang21}), this is the only sample of YSGs with both temperature and luminosity measurements. After excluding stars with $\lum < 4$ to avoid contamination by lower-mass evolved stars (see, for example, \citealt{levesque17}), we cross-matched this sample to the latest version of the \tess~Input Catalog (TIC, \citealt{stassun18}) available on the Mikulski Archive for Space Telescopes (MAST), and selected all stars with a \tess~magnitude fainter than $T=4$ (above which \tess~begins to saturate), and brighter than $T=12$ (i.e., bright enough to detect sub-ppt-level variability in YSGs, see \citealt{dornwallenstein19,dornwallenstein20b}).

We selected all stars that had been observed by \tess~at two-minute cadence in the southern hemisphere using target lists for \tess~Sectors 1-13 and 27-39.\footnote{\tess~target lists are available online at \url{https://tess.mit.edu/observations/target-lists/}} This results in a total of 219 stars. However, \citet{neugent10,neugent12_ysg} determined membership in the Magellanic Clouds via radial velocities. In the intervening years, the \Gaia~mission \citep{gaia16} has provided us with extremely high-quality astrometry which can be used to further refine our sample. We cross-matched our sample with the catalog of Magellanic Cloud stars from \citet{gaiacollab18}, which uses data from \Gaia~DR2 \citep{gaia18}. This allows us to discard a further 18 stars in our sample as being likely foreground objects. We list the names, TIC numbers, and \Gaia~DR2 parallaxes and uncertainties (to compare with \citealt{gaiacollab18}) for LMC stars discarded at this step in Table \ref{tab:tossed}. The only LMC star in this list with a spectral type available in the literature is HD 268943 (evolved A0 supergiant; \citealt{cannon36}). Three SMC stars have literature spectral types --- UCAC2 1077812 (G6Iab; \citealt{gonzalezfernandez15}), Flo 739 (F7I; \citealt{florsch72}), and UCAC2 1249286 (G5I; \citealt{neugent10}) --- all of which appear to have been assigned an incorrect luminosity class. All stars have parallax values of-order 0.1 mas, well-above the median LMC/SMC parallax of -19/-0.9 $\mu$as respectively from \citet{gaiacollab18}, confirming that these stars are all likely foreground objects. 

After discarding these objects, the sample contains a total of 201 stars. However, upon inspection of the spectral types available in the literature for these stars, we found that 75 of them are actually B-type stars. While \citeauthor{neugent10} did deliberately include these stars in their sample, they were not listed as B supergiants at the time, and, because of the near-infrared excess in these stars caused by free-free emission in their winds, the derived temperatures from $J-K$ photometry were sensible for AFG supergiants. While \tess~is sensitive to a number of interesting physical processes in B supergiants, none of them are the focus of this work, and so we discard all such objects. This leaves us with a sample of 126 stars --- 101 in the LMC, and 25 in the SMC --- which nearly doubles the size of the sample in \citet{dornwallenstein20b}, and extends our work to the SMC for the first time.

\begin{deluxetable}{lccc}
\tabletypesize{\scriptsize}
\tablecaption{Names, TIC numbers, parallaxes, and uncertainties for the stars discarded after cross-matching with \citet{gaiacollab18}.\label{tab:tossed}}
\tablehead{\colhead{Common Name} & \colhead{TIC Number}  & \colhead{$\varpi$} & \colhead{$\sigma_\varpi$}\\
\colhead{} & \colhead{}  & \colhead{[mas]} & \colhead{[mas]} } 
\startdata
\cutinhead{LMC}
2MASS J05172733-6902538 & 179210681 & 0.1557 & 0.0310 \\ 
MACHO   3.6848.668 & 179636847 & 0.2390 & 0.0223 \\ 
2MASS J04521129-7148341 & 30035351 & 0.4251 & 0.0192 \\ 
2MASS J05022839-7209032 & 140831368 & 0.0548 & 0.0257 \\ 
2MASS J04462599-6950186 & 294868316 & 0.0921 & 0.0208 \\ 
2MASS J04530398-6937285 & 30032006 & 0.0850 & 0.0273 \\ 
MACHO   8.9024.14 & 277028154 & 0.3320 & 0.0270 \\ 
2MASS J04463462-6704279 & 294872402 & 0.4372 & 0.0348 \\ 
HD 268943 & 30931288 & 0.2189 & 0.0491 \\ 
\cutinhead{SMC} 
UCAC2    791862 & 182517311 & 0.3403 & 0.0206 \\ 
UCAC2   1077812 & 267496747 & 0.2534 & 0.0259 \\ 
Flo 739 & 183799843 & 0.5495 & 0.0207 \\ 
UCAC2   1249286 & 52014238 & 0.1500 & 0.0242 \\ 
{[M2002]} SMC  73541 & 183306212 & 0.6740 & 0.0244 \\ 
Gaia DR2 4686397459879956992 & 426012845 & 0.6944 & 0.0295 \\ 
{[M2002]} SMC  76389 & 183495292 & 0.5260 & 0.0213 \\ 
UCAC2    856760 & 182734686 & 0.3906 & 0.0265 \\ 
Gaia DR2 4686415425725954944 & 426012941 & 0.2676 & 0.0262 \\ 
\enddata
\end{deluxetable}

Using the Python package \texttt{astroquery}, we then queried MAST and downloaded all available two-minute cadence lightcurves for each target. As in \citet{dornwallenstein20b}, we used the \texttt{PDCSAP\_FLUX} lightcurves that have been corrected for systematic trends, and stitch together lightcurves from different \tess~sectors\footnote{Each \tess~sector is the length of two \tess~orbits around Earth, approximately 27 days.} by dividing each sector's data by the median flux in each sector. Due to the location of the LMC in the \tess~southern continuous viewing zone, many of the LMC stars in our sample were observed for approximately a year during either the first or third year of \tess~operations, or for upwards of two years in some cases with a year-long gap in observations during which \tess~observed the northern ecliptic hemisphere. Even the least observed LMC stars still have \tess~lightcurves spanning $\sim$190 days (seven sectors) of observations. Because the SMC was in the \tess~field of view for less time, SMC stars have at most $\sim$55 days (two sectors) of \tess~observations.



\begin{figure}[t!]
\includegraphics[width=0.48\textwidth]{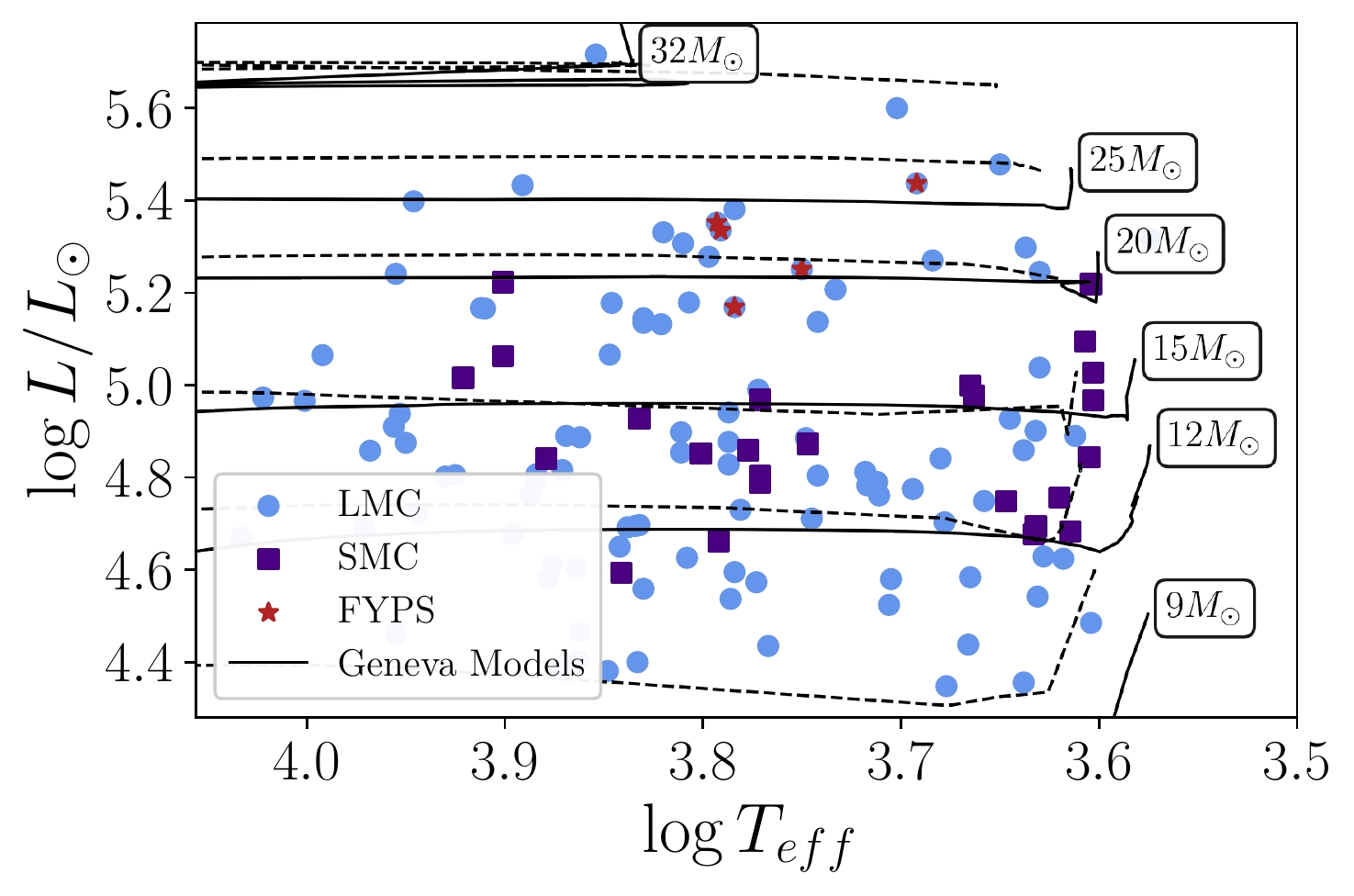}
\caption{Hertzprung-Russel diagram showing the effective temperatures and luminosities of the 126 stars in our sample. Stars in the LMC are shown as blue circles, and stars in the SMC as purple squares. The five FYPS identified in \citet{dornwallenstein20b} are shown as red stars. Black lines show rotating evolution models at LMC metallicity (solid line, \citealt{eggenberger21}) and SMC metallicity (dashed line; \citealt{georgy13}), with their initial masses indicated.}\label{fig:sample_hr}
\end{figure}

Figure \ref{fig:sample_hr} shows the distribution of stars in our sample in the Hertzprung-Russell (HR) diagram. Stars located in the LMC are shown as blue circles, while stars in the SMC are indicated with purple squares. Red stars show the location of the five LMC stars identified as FYPS by \citet{dornwallenstein20b}. Evolutionary tracks from the Geneva group are shown as black lines with their initial masses indicated. Solid lines correspond to rotating $Z=0.006$ (LMC metallicity) tracks from \citet{eggenberger21}, while dashed lines are rotating $Z=0.002$ (SMC metallicity) tracks from \citet{georgy13}. We note that these tracks incorporate ``standard'' recipes for RSG mass loss, and that all tracks shown finish their evolution as RSGs. This includes the 32 $M_\odot$ $Z=0.006$ track, which undergoes blueward evolution while losing $\sim$half of its envelope mass, but then evolves redward during the final few thousand years of its life.

\subsection{Identifying FYPS}\label{subsec:identify}

Throughout their lifetimes, massive stars display stochastic low frequency (SLF) variability. This variability is manifested in the periodograms of massive star lightcurves as a rising trend towards lower frequencies, similar in morphology to the signature of granulation in sun-like and red giant stars. SLF variability was discovered in OB stars by \citet{blomme11}, and \citet{bowman19} demonstrated that it is a ubiquitous trait of hot stars. This SLF variability has been attributed either to internal gravity waves (IGW; \citealt{bowman19b,bowman20}), subsurface convection \citep{cantiello21}, or wind-driven processes \citep{krticka21}. SLF variability in cool supergiants is a recent discovery \citep{dornwallenstein20b}, and its driving mechanism (and whether or not it is linked with the SLF variability seen in hot stars) is still unknown. However, it is found in all of the stars that we consider in this work, is significant over the instrumental background across the entire frequency range where FYPS pulsations appear, and has an amplitude comparable with the pulsations exhibited by FYPS.

This means that identifying FYPS and measuring their oscillation frequencies with high precision is a unique challenge compared to other varieties of pulsators in which the oscillations have higher amplitudes, the background is uncorrelated (i.e., white noise), or the correlation timescale of the background is significantly different than the pulsation periods of interest (e.g., many solar-like oscillators). In \citet{dornwallenstein20b}, we first fit the amplitude spectrum (calculated as the square root of the periodograms derived from the \tess~lightcurve) for each star in our sample with a quasi-Lorentzian function:
\begin{equation}\label{eq:rednoise}
    \alpha(f) = \frac{\alpha_0}{1+(2\pi\tau_{\rm char} f)^\gamma}+\alpha_w.
\end{equation}
Here $f$ is the frequency, $\alpha_0$ is the amplitude as $f\rightarrow0$, $\tau_{\rm char}$ is a characteristic timescale on which the noise is correlated, $\gamma$ sets the slope of the red noise, and $\alpha_w$ corresponds to the white noise floor at the highest frequencies. This is identical to the function adopted by \citet{bowman19}, who instead use a characteristic frequency, $\nu_{char}=(2\pi\tau_{\rm char})^{-1}$. After fitting the amplitude spectra, we divided the periodogram by the square of the best-fit function in order to identify strong peaks in the residual power spectrum. We then performed an prewhitening procedure, modelling the lightcurve as a sum of sinusoids with frequencies inferred from the residual power spectrum, iteratively adding periodic components to the model until a minimum in the Bayesian Information Content (BIC, \citealt{schwarz78}) was reached. If a model with at least one frequency was preferred (i.e., had a lower BIC) over a constant model, then we inferred that the recovered frequencies are actually present in the data.

\begin{figure*}[t!]
\plotone{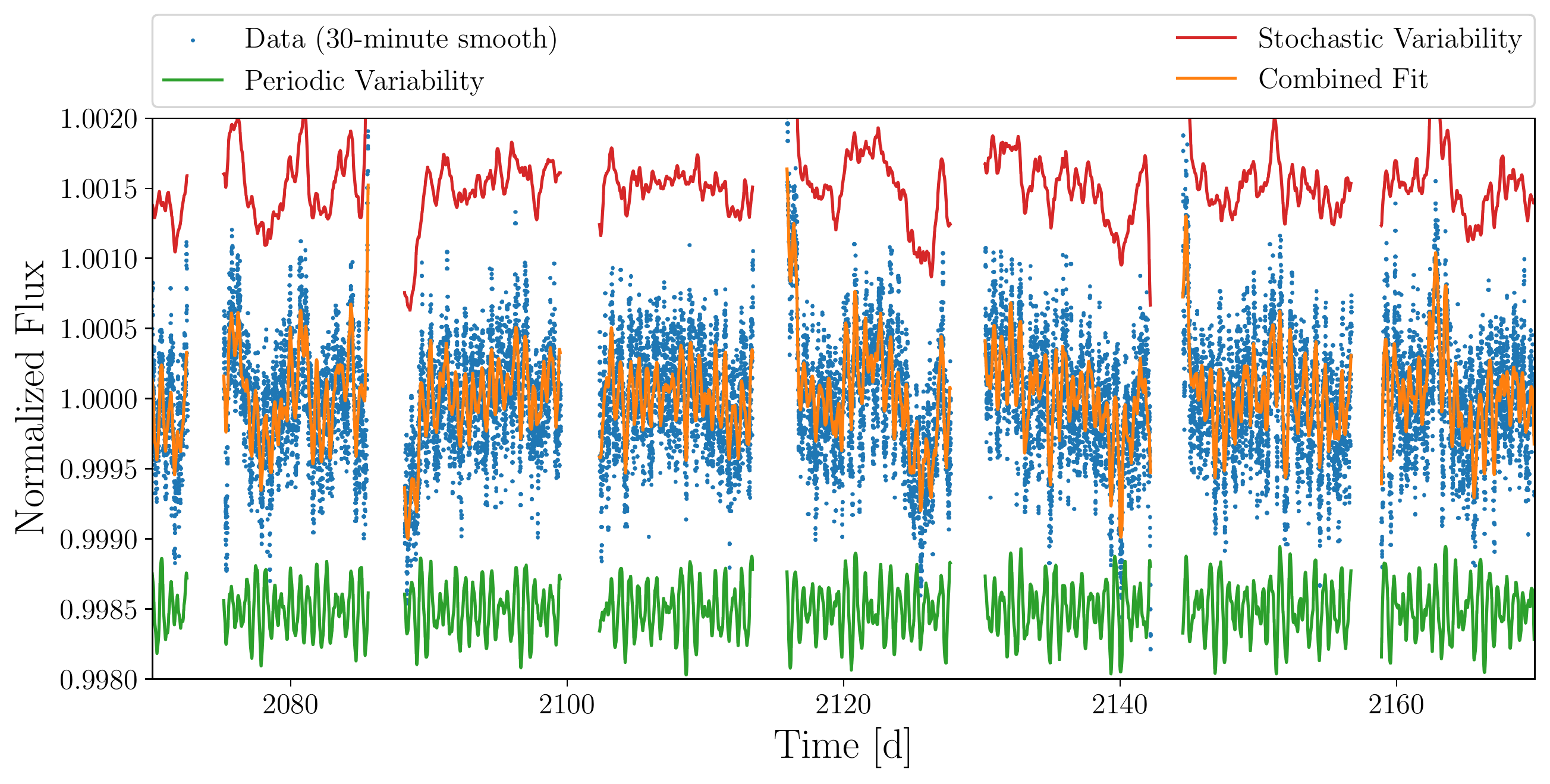}
\caption{Example of the Gaussian process-based approach to identifying FYPS. The data in blue is a section of the \tess~lightcurve of the FYPS HD 269953, with a 30-minute Gaussian smoothing applied for clarity. The orange line shows the maximum-likelihood GP fit. The individual components are shown in green (periodic mean function) and red (SLF variability), with offsets applied for clarity.} \label{fig:GP_deblend}
\end{figure*}

While this method was suitable for discovering FYPS, this method used a $\chi^2$-like metric to assess the likelihood function, which assumes that each data point is independent. Due to the SLF variability in our sample, this is not true; the data are correlated by definition! Furthermore, we are concerned that the pulsation frequencies (and associated uncertainties) measured with our prewhitening method could have been biased by the SLF variability, which we did not account for.

For these reasons, we developed a method that uses Gaussian processes (GPs; see \citealt{rasmussen06} for a comprehensive overview) to simultaneously model the coherent and SLF variability in the time domain. GPs are a flexible class of methods that is capable of characterizing the SLF variability parameters rapidly and with high precision, even in the limit where instrumental white noise is significant \citep{bowman22}. Following the notation of \citet{foremanmackey17}, the GP log-likelihood function can be written (to within a constant) as:
\begin{equation}\label{eq:GP_ll}
    \log \mathcal{L}(\pmb{\theta}, \pmb{\alpha}) = -\frac{1}{2}[\pmb{y}-\pmb{\mu_\theta}]^{\rm T} K_{\pmb{\alpha}}^{-1}[\pmb{y}-\pmb{\mu_\theta}] -\frac{1}{2}\log \det K_{\pmb{\alpha}}
\end{equation}
where $K_{\pmb{\alpha}}$ is the covariance matrix of the stochastic variability (whose components we assume can be modelled by some function with hyperparameter vector $\pmb{\alpha}$), and $[\pmb{y}-\pmb{\mu_\theta}]$ is a vector of residuals between the data $\mathbf{y}$ and a model for the mean flux, $\pmb{\mu_\theta}$ with parameter vector $\pmb{\theta}$ whose entries we are interested in. By optimizing this likelihood simultaneously with respect to $\pmb{\alpha}$ and $\pmb{\theta}$, we account for any covariances between the two sets of parameters that would otherwise be ignored if we simply prewhitened the data before characterizing the SLF variability. 

The flexibility in the choice of $\pmb{\mu_\theta}$ in Equation \eqref{eq:GP_ll} allows us to cast the task of identifying pulsating stars as a model selection problem: is a given lightcurve modelled better as SLF variability plus some sum of sinusoids, or as just SLF variability superimposed on a constant flux? To determine which model best fits the observations with the minimum number of free parameters, we calculate the BIC:
\begin{equation}
    {\rm BIC} = -2\log \mathcal{L} + m\ln N
\end{equation}
where $m$ is the number of free parameters, and $N$ is the number of observations in the lightcurve. To calculate the GP log-likelihood, we use {\sc celerite2} \citep{foremanmackey17,foremanmackey18}, and model the SLF variability using the {\tt SHOTerm} kernel following \citet{bowman22}. This kernel has also been used successfully to model granulation in sun-like stars \citep[e.g.][]{pereira19}. To maximize the log-likelihood, we use {\sc PyMC3} \citep{salvatier16}, which is compatible with {\sc celerite2}, and adopt either a constant mean function, or some sum of sinusoids with trial frequencies, amplitudes, and phases extracted from the lightcurve via the iterative prewhitening procedure from \citet{dornwallenstein20b}. We describe the various steps that we use in detail in Appendix \ref{app:GP}. Figure \ref{fig:GP_deblend} shows an example of this process applied to the FYPS HD 269953. The data are a 100-day cutout of the \tess~lightcurve with a thirty-minute Gaussian smooth applied. The orange shows the maximum likelihood GP fit with the minimum BIC model, and the red/green lines show the stochastic/periodic components of the fit, with offsets applied for clarity. The mean function contains 10 independent frequencies.


This procedure recovers a total of 91 stars that exhibit periodic variability. For the majority of our sample, the coherent variability is at relatively high frequencies ($0.2-10$ d$^{-1}$); as discussed in \citet{dornwallenstein19,dornwallenstein20b}, these timescales are too fast to be attributed to surface rotation or orbital modulation in a binary system, making pulsations the most-likely culprit.

\subsection{Removing Contaminants}

\subsubsection{Contamination by Nearby Variable Stars}\label{subsubsec:ogle}

All of our targets are located in the Magellanic Clouds, and each \tess~pixel is 21'' on a side ($\sim$17 ly at the distance of the LMC). Therefore, crowding and contamination are significant issues. In \citet{dornwallenstein20b} we used a statistical argument to show that the clustering in the HR diagram of the five stars we identified as FYPS ruled out contamination either by a pulsating binary companion or by a nearby pulsating star in the \tess~aperture. Our current sample is much larger, and we now suspiciously find ``pulsating'' stars throughout the HR diagram, a fact that leads us to conclude that contamination is likely affecting our sample. 

To directly assess the impact of contamination from nearby variable stars on a star-by-star and frequency-by-frequency basis, we turned to the publicly available archive of the Optical Gravitational Lensing Experiment (OGLE; \citealt{udalski15}). For each star, we query the OGLE catalog\footnote{\url{https://ogledb.astrouw.edu.pl/~ogle/OCVS/catalog_query.php}} for all stars brighter than $I=17$ within 150'', and all OGLE periods shorter than 15 days. For each frequency recovered from the \tess~data ($f_{\rm TESS}$) and each frequency observed in a nearby OGLE lightcurve ($f_{\rm OGLE}$), we can compute the fractional difference in frequency, 
\begin{equation}
    \delta f = \frac{|f_{\rm TESS} - a f_{\rm OGLE}|}{f_{\rm TESS}}
\end{equation}
where $a$ is either an integer or the inverse of an integer. We then consider the probability that a frequency drawn randomly from a uniform distribution spanning $\Delta\log{f}=4$ orders of magnitude in log-space (approximately the range of frequencies that we recover) would be found within $\delta f$ of a frequency found within the set of OGLE frequencies (i.e., the probability of a false positive), which is
\begin{equation}\label{eq:pfalse}
    P_{\rm false} = \frac{2 N N_a \delta f}{\Delta\log f\ln 10}
\end{equation}
where $N$ is the number of OGLE stars considered and $N_a$ is the number of values for $a$ that we consider (six in total, using $a = $ 1, 2, 4, 1/2, 1/3, and 1/4). We also consider combinations of two OGLE frequencies (i.e., $f_{\rm OGLE} = f_{\rm OGLE, 1} \pm f_{\rm OGLE, 2}$), in which case the factor of $N$ in Eq. \eqref{eq:pfalse} becomes $N(N-1)$.\footnote{There are $N(N-1)/2$ combinations of two stars, and factor of 1/2 is cancelled by testing both the sum and difference of the two OGLE frequencies.} We then reject all frequencies with $P_{\rm false} < 10\%$. After performing this process, we are left with a total of 57 stars with at least one frequency remaining.

\begin{figure}[ht!]
\plotone{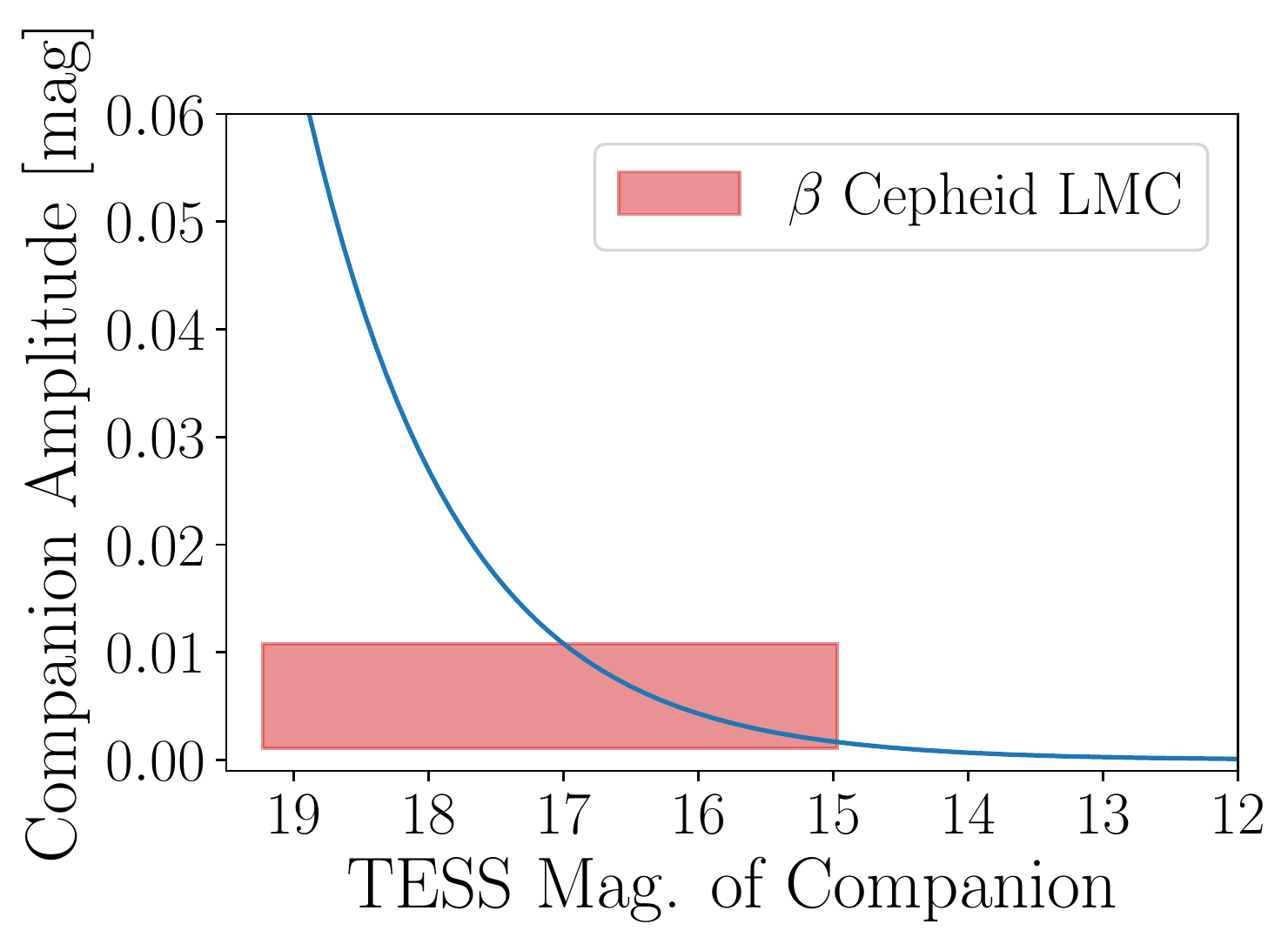}
\caption{Amplitude in magnitudes of a companion star as a function of the companion star's \tess~magnitude that would appear as 100 ppm variability when blended with a non-variable $T=12$ primary. The red box corresponds to the typical brightness and amplitude range of a $\beta$ Cepheid star in the LMC. The blue line intersects the red box, which indicates that the observed variability in this toy example is consistent with contamination by a pulsating B type companion.} \label{fig:bcep_contaminant}
\end{figure}

\subsubsection{Contamination by Pulsating Binary Companions}

We next turn our attention to the possibility that the lightcurves have been contaminated by a pulsating binary companion. Massive stars are preferentially born into binary systems \citep{sana12,duchene13,sana14,eldridge17}, and from evolutionary modelling, the most likely companion of an evolved supergiant is likely to be a B-type stars (or possibly a late O star), as less massive companions are unable to form fast enough to reach the main sequence by the time a massive star completes its evolution \citep[e.g.][]{neugent18b,neugent19}.\footnote{Due to the fact that a star only spends a small fraction of its life as a post-main sequence object, a binary system composed of two evolved supergiants is exceedingly unlikely. If any of these stars were actually in such a system, we would expect their Hydra spectra from \citet{neugent12_ysg} to show signs of binarity, which they do not.} Both $\beta$ Cepheid variables of spectral types O and B as well as Slowly Pulsating B (SPB) stars exhibit pulsations on similar timescales as we observe in our sample with amplitudes ranging from approximately 1-10 ppt \citep{balona20}. 

To deduce which lightcurves are potentially contaminated by a B-type companion, we can pose the following thought experiment. Say we observe a star with magnitude $T$ that appears to vary with amplitude $Y$ (or, in magnitude units, $2.5\log{1+Y}$). Imagine this star is actually constant, and bright enough that it is the dominant source of light in the \tess~bandpass, but is contaminated by a companion that is $\Delta T$ magnitudes fainter (corresponding to a flux ratio of $R = 10^{-\Delta T/2.5}$), pulsating at an intrinsic amplitude $X > Y$. In order for us to mistakenly determine that the primary star is variable with an amplitude $Y$, the true amplitude must be $X=Y/R$, (or in magnitude units, $-2.5 \log (1+Y/R)$). Figure \ref{fig:bcep_contaminant} shows a toy example, in which a $T=12$ primary is contaminated by a secondary whose \tess~magnitude is shown on the x-axis. This imaginary system is observed to vary with an amplitude of 100 ppm (0.1 mmag); the blue line shows the intrinsic amplitude of the companion that would result in the observed amplitude. 

We can compare this result with the typical \tess~magnitude and pulsation amplitude of a $\beta$ Cepheid variable. To do this, we use {\sc PySynphot} \citep{pysynphot13}, along with synthetic stellar spectra from \citet{kurucz93} and the publicly-available \tess~bandpass\footnote{\url{https://heasarc.gsfc.nasa.gov/docs/tess/data/tess-response-function-v2.0.csv}} to calculate synthetic \tess~photometry of a B0V and B9V star in both the SMC and LMC, using the parameters for these stars from \citet{silaj14}, distance moduli from \citet{kovacs00a,kovacs00}, and typical extinction for the Clouds from \citet{gordon03}, assuming $R_V = 2.74/3.41$ in the SMC/LMC respectively. Typical $\beta$ Cepheid amplitudes in the \tess~bandpass are taken from \citet{balona20}. The red box in Figure \ref{fig:bcep_contaminant} shows the result of our synthetic photometry, bounding the region in this parameter space in which a typical pulsating B star in the LMC resides. The blue line intersects with the red box, indicating that in this toy example, the observed variability is consistent with the star's lightcurve being contaminated by a $\beta$ Cepheid companion. We perform this process for every remaining star in our sample with at least one frequency, using the star's \tess~magnitude to determine whether the star's variability could be due to contamination from a pulsating B-type companion. We use the properties of a B-type star in the LMC or SMC depending on the star's host galaxy, and remove a star if all of its frequencies are consistent with such contamination (21 stars in total). We note that this argument also applies in the case of chance spatial alignment between a star in our sample and a single B star in its host galaxy.

\section{Results}\label{sec:results}

\subsection{Pulsating Stars in the Upper HR Diagram}\label{subsec:fpulse}

From the list of the 91 pulsating stars recovered by the GP procedure described above, we discard all frequencies that are likely to be contaminants from nearby stars, and all objects whose lightcurves could be contaminated by a pulsating B star (21 objects). This results in a sample of 36 {\it bona fide} pulsators. Using the luminosity and temperature estimates from \citet{neugent10,neugent12_ysg}, we now wish to visualize the location of these pulsating stars in the HR diagram.

To do this, we use kernel density estimation, a technique that replaces each point in the HR diagram with a two-dimensional Gaussian kernel of a given width to estimate the distribution of objects in the HR diagram. Because the dynamic range of the luminosity and temperature measurements are slightly different, and most implementations of kernel density estimates (KDEs) use a symmetric kernel for multi-dimensional density estimation (i.e., the kernel width is the same in all dimensions), we first use {\sc Scikit Learn} \citep[][]{scikit-learn11} to scale the data such that the transformed $\teff$ and $\lum$ measurements each have a mean of 0, and a standard deviation of 1. We then use {\sc Scikit Learn} to fit the transformed $\lum$ and $\teff$ measurements of the entire sample using a KDE with a kernel size of 0.5 (corresponding to half the standard deviation of the sample in $\teff$ and $\lum$ when transformed back into the observed HR diagram). We also fit just the sub-sample of 58 confirmed pulsating stars. We compute both density estimates on a grid of 100x100 points in the transformed variables, before transforming this grid back into the observed HR diagram. From these results, we can compute the fraction of stars that pulsate, $f_{\rm pulse}$, as a function of position in the HR diagram as:
\begin{equation}\label{eq:fpulse}
    f_{\rm pulse}(\teff,\lum) = \frac{KDE_{\rm pulse}}{KDE_{\rm all}}\frac{N_{\rm pulse}}{N_{\rm all}}
\end{equation}
where the subscripts ``pulse'' and ``all'' correspond to the sample of pulsators and the entire sample, and $N$ is the number of stars in each subset.

\begin{figure*}[ht!]
\includegraphics[width=\textwidth]{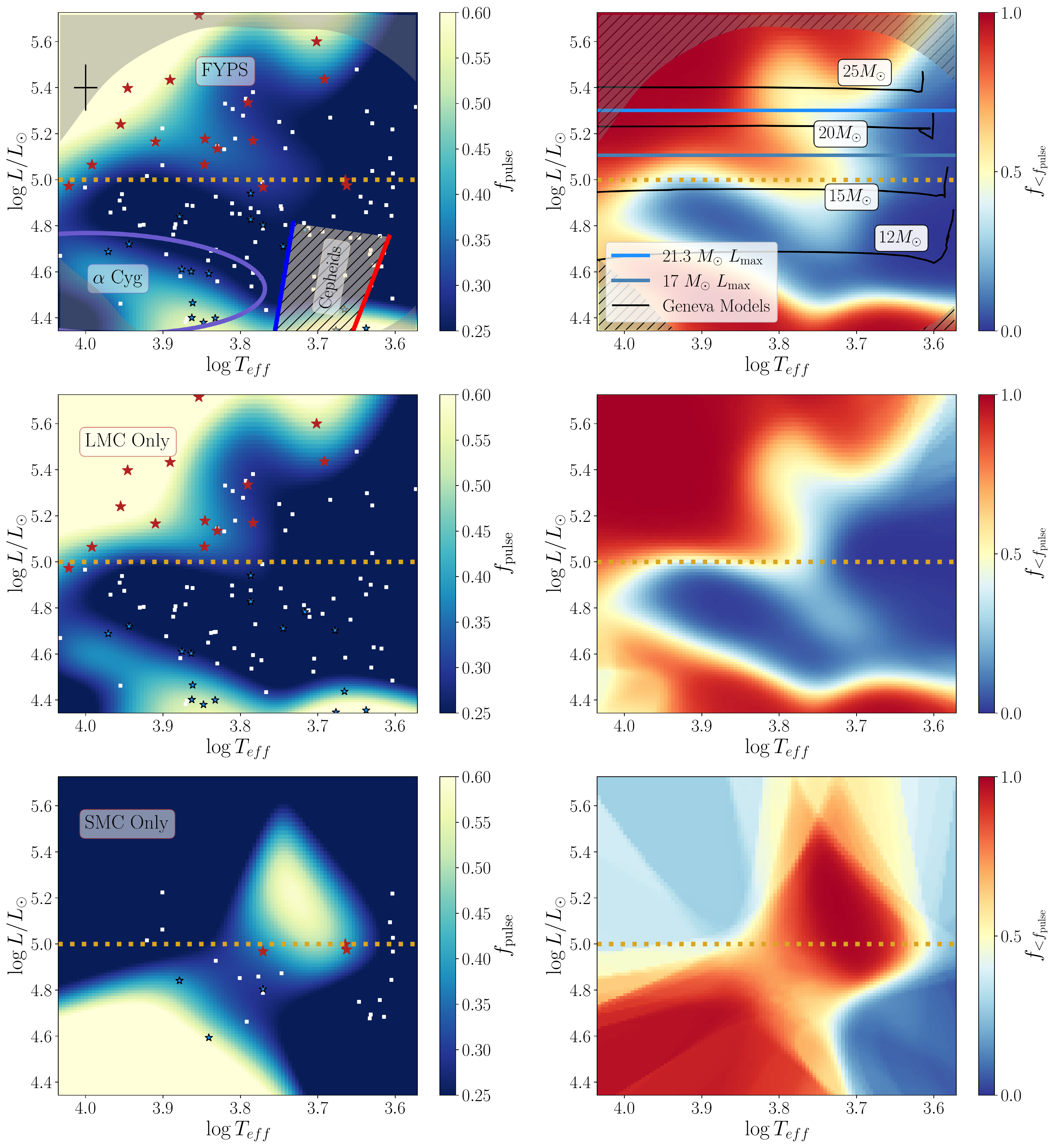}
\caption{{\it Left panels}: The pulsator fraction $f_{\rm pulse}$ shown as a function of location in the HR diagram, with $f_{\rm pulse}$ increasing from blue to yellow. The stars in our sample are shown as white squares, with confirmed pulsators shown as blue stars with black outlines, FYPS (pulsators with $\lum \geq 5.0$) shown as red stars, and the original five FYPS discovered by \citet{dornwallenstein20b} shown as red stars with black outlines. The typical uncertainty in $\teff$ and $\lum$ is shown as the black errorbar in the top left corner, and the region where our sample is incomplete is shown as the dark grey outline. The $\lum \geq 5.0$ is shown as a dotted goldenrod line, the Cepheid instability strip is shown as a hatched, shaded region, and the region containing low-luminosity $\alpha$ Cygni variables is indicated by the periwinkle ellipse. {\it Right panels}: An estimate of the statistical significance of the structures seen in the left panels. The fraction of simulated $f_{\rm pulse}$ values drawn from 10,000 bootstrap simulations whose value is less than the true value of $f_{\rm pulse}$, $f_{< f_{\rm pulse}}$, with $f_{< f_{\rm pulse}}$ increasing from red to blue. Regions in dark blue/red are statistically significant under/overdensities of pulsators. For comparison, we overplot $Z=0.006$ evolutionary tracks from \citet{eggenberger21}, the minimum FYPS luminosity at $\lum \geq 5.0$ in goldenrod, and the maximum luminosity of a 17 and 21.3 $M_\odot$ star calculated using the relationship presented in \citet{kochanek20} in dark and light blue respectively. The top panels show the results of this computation for the entire sample, with the middle and bottom panels showing the results restricted to the stars in the LMC and SMC, respectively.} \label{fig:kde_fpulse_bootstrap}
\end{figure*}

The top-left panel of Figure \ref{fig:kde_fpulse_bootstrap} shows the result of this calculation. The value of $f_{\rm pulse}$ is indicated by the color, with blue corresponding to lower values and yellow corresponding to higher values. To highlight the range of values in the plot, we limit the colorbar to correspond to the range $0.25 \leq f_{\rm pulse} \leq \mathbf{0.6}$. The grey shaded region in the outer boundary of the plot corresponds to where $KDE_{\rm all} < 0.015$ as an {\it ad hoc} way of showing where the density of stars in the HR diagram is too low to make meaningful interpretations. The non-pulsating stars in our sample are shown as small white squares, and pulsating stars are shown as slightly larger blue stars with black outlines. The typical uncertainties on $\teff$ and $\lum$ are shown as an error bar in the top left corner of the plot. The figure itself is quite detailed, so we pause here to note a critical takeaway: we detect a high fraction of pulsators at luminosities above $\lum=5.0$.

Now, examining the figure in detail, in the lower-right we find a pronounced paucity of pulsators. For comparison, we show the theoretical Cepheid instability strip computed at $Z=0.006$ (i.e., the metallicity of the LMC) from \citet{anderson16}, which we plot as a cross-hatched shaded region with the cool/warm edge of the instability strip indicated in red/blue respectively. We infer that 
\begin{enumerate}
    \item Our sample is largely insensitive to Cepheid variables, or contamination by nearby Cepheids (which we addressed in \S\ref{subsubsec:ogle}). This is unsurprising as typical Cepheid variability occurs at high amplitudes on long timescales, which we expect to be mostly removed by the \tess~SPOC processing of the {\tt PDCSAP} lightcurves. The exception to this is the three lowest luminosity stars in this quadrant of the figure; at this luminosity, \tess~does become sensitive to the typical periods for Cepheids in the LMC/SMC \citep[e.g.][]{soszynski18}.
    \item The variability that we detect really does occur at the observed frequencies; we are not seeing low-frequency variability that is modulated into a higher frequency band by systematic or instrumental effects. Otherwise, we would expect to recover luminous Cepheids, which we don't.
\end{enumerate}

The remainder of the plot shows two clumps of high $f_{\rm pulse}$; one occurs at high temperature ($\teff \gtrsim 3.8$) and low luminosity ($\lum \lesssim \mathbf{4.8}$), and the other above $\lum \gtrsim 5.0$, especially for temperatures above $\teff\lesssim\mathbf{3.7}$. Between the terminal age main sequence (TAMS) and the red supergiant phase, massive stars evolve across the HR diagram at approximately constant luminosity. As a result, the separation that we show as a dotted goldenrod line at $\lum=5.0$ corresponds to a boundary in {\it initial mass}. This boundary itself is quite interesting, as it roughly corresponds to stars with initial masses of $\sim$18-20 $M_\odot$, and marks the transition to significantly higher RSG mass loss rates observed by \citet{humphreys20}. For this reason, we follow our previous work and associate the lower luminosity clump (which we circle in Figure \ref{fig:kde_fpulse_bootstrap} with a periwinkle ellipse) with the $\alpha$ Cygni variables. These pulsating B and A supergiants (blue supergiants, BSGs) are thought to be post-RSG objects \citep[][]{saio13,georgy21}. \citet{georgy21} compare their models with a number of $\alpha$ Cyg variables that span a wide range in luminosity, including relatively low-luminosity stars down to $\lum \geq 4.5$; we also recover $\alpha$ Cyg variables at comparably low luminosity. Our work suggests a division between these lower luminosity objects and the higher luminosity stars that we discuss further below. While estimates on the minimum-mass star that should experience a post-RSG phase through single-star evolution vary, it would be quite surprising to find post-RSGs at such low luminosity. These lower luminosity $\alpha$ Cyg variables may therefore be the result of binary interactions. Quite interestingly, two studies of the SN Ib PTF13bvn identify the progenitor as a stripped star in a binary system whose properties are consistent with the $\alpha$ Cyg stars in our sample \citep{bersten14,eldridge15}. As these stars are not the focus of the present work, we leave further speculation to other authors.

We associate the high-luminosity pulsators with the recently-discovered Fast Yellow Pulsating Supergiants (FYPS), and we highlight the thirteen pulsating stars in this part of the HR diagram with red star-shaped markers. We pause here to note two important details. First, of the five stars identified as FYPS by \citet{dornwallenstein20b}, only three of them are identified as such here (HD 269953, HD 268687, and HD 269840). All of the frequencies recovered from the other two stars, HD 269110 and HD 269902, were rejected as being contaminants from nearby stars. Second, there are a number of stars immediately below the goldenrod line in the figure. Indeed, if we lower the threshold to $\lum=4.95$ (i.e., approximately half of the typical error on the luminosity measurement), an additional four pulsating stars are selected. We consider it reasonably likely that these stars are associated with the higher luminosity pulsators, and therefore include them in the discussion of FYPS going forward (while continuing to refer to the boundary at $\lum=5$). However, we note that further work is needed to determine exactly where the minimum FYPS luminosity boundary should be located.

All told, we identify 17 stars as FYPS, increasing the number of known FYPS by almost a factor of four. We tabulate all of the stars identified as FYPS in Tables \ref{tab:lmc_fyps} and \ref{tab:smc_fyps}, and show their \tess~lightcurves and periodograms (after normalizing by the shape of the SLF variability) in Figures \ref{fig:lmc_fyps} and \ref{fig:smc_fyps}. In these figures, we indicate which frequencies are attributable to nearby variables, and which we attribute to the star itself. Pulsating stars not identified as FYPS are listed in Tables \ref{tab:lmc_acyg} and \ref{tab:smc_acyg}, and the remaining nonpulsating stars at all luminosities are listed in Tables \ref{tab:lmc_np} and \ref{tab:smc_np}. 

We now return to the increase in $f_{\rm pulse}$ at high luminosities around $\teff \approx 3.8-3.9$ ($\sim$6300-7950 K). This temperature regime corresponds to the transition between stellar atmospheres with electron-scattering as the main source of opacity, and those where H$^{-}$ opacity dominates --- and where efficient surface convection begins to develop.\footnote{Similar physics are responsible for the ``Kraft break,'' a transition in the observed rotation rates in low mass main sequence dwarfs across this temperature boundary \citep{kraft67}.} Therefore, we expect a transition in the pulsational properties of FYPS at this temperature. Indeed, that is precisely what we observe, as we discuss in \S\ref{subsec:astero}. However, we note that the stars we identify as FYPS are fairly evenly-distributed across the upper HR diagram and the transition in pulsational properties that we discuss below is a smooth one. Therefore, this may just be a result of how well the HR diagram is sampled in this region.

Finally, the middle-left and bottom-left panels of Figure \ref{fig:kde_fpulse_bootstrap} show the results of computing $f_{\rm pulse}$ exclusively on the stars residing in the LMC and SMC, respectively. These panels largely illustrate the effects of our sampling --- e.g., most of the signal seen in the upper-left panel is driven by the behavior of stars in the LMC --- and the limitations of the KDE when extrapolated beyond the sample coverage, especially in the SMC. However, the key features of the plot --- a transition point at $\lum=5.0$, and the lack of low-temperature, low-luminosity pulsators exist irrespective of the host galaxy. Interestingly, the high luminosity pulsators in the SMC appear to be found preferentially at lower temperatures, but with so few stars, we cannot make any conclusions about possible metallicity effects.


\subsubsection{A Note on \texorpdfstring{$\alpha$}{} Cygni Variables \texorpdfstring{\&}{} Nomenclature}

It is now apparent that the ``Y'' (for yellow) in FYPS is now somewhat inaccurate; FYPS are found among both FG supergiants and higher-temperature A supergiants as well. As a result, there is now considerable overlap between the properties of the warmer FYPS and the properties of $\alpha$ Cygni variables. Indeed, measurements from \citet{schiller08} place Deneb, the $\alpha$ Cygni prototype, squarely in the upper left quadrant of the left panel of Figure \ref{fig:kde_fpulse_bootstrap}. Of course, the precise definition of an $\alpha$ Cygni variable in the literature is somewhat unclear; early catalogs of variable stars define $\alpha$ Cygni variables as luminous B and A supergiants that display $\sim$0.1 mag variability \citep{kholopov85}. That definition then grew to encompass a broad swath of supergiants of nearly all spectral types, on the assumption that these stars form a smooth evolutionary sequence \citep[e.g.][]{vangenderen89,vangenderen92}. This includes all manner of variable objects, including luminous blue variables and B[e] supergiants \citep[e.g.][]{vangenderen99,vangenderen02}, whose variability may be caused by a plethora of physical effects. More recent work has winnowed this definition back to just B and A supergiants \citep[e.g.][]{samus17}, which pulsate nonradially on periods of a few days or longer. While the pulsations we observe in the \tess~data are typically faster and have lower amplitudes, this may be due to the fact that ground-based surveys just aren't sensitive to the short periods and low amplitudes that \tess~is. 

So are the warmer FYPS in our sample truly FYPS, or are they $\alpha$ Cygni variables? Should the term ``FYPS'' be used solely to refer to the pulsating FG supergiants in our sample? Unfortunately, very little is known about pulsations in both FYPS and $\alpha$ Cygni variables from a theoretical standpoint, making it difficult to determine the connection between the two; both classes of objects are thought to be candidate post-RSG objects \citep{saio13,dornwallenstein20b,georgy21}. However, as we mention above, some $\alpha$ Cygni variables have luminosities below $\lum=5.0$ \citep{vangenderen89} and thus should not experience post-RSG evolution. As we discuss further in \S\ref{subsec:evostatus}, previous work has shown that a small number of the most luminous stars we identify as FYPS in our sample are also likely pre-RSG objects. Furthermore, while there is a decrease in $f_{\rm pulse}$ at intermediate temperatures around $\teff\approx3.8$, six of the recovered FYPS reside in this decrease, and we observe a continuum in the behavior of the recovered pulsation frequencies across this region (\S\ref{subsec:astero}), raising doubts that the warm (A type) and cool (FG type) FYPS form distinct classes.

 Ultimately, while more precise nomenclature is absolutely necessary, introducing additional nomenclature at this juncture is likely to result in further confusion until our theoretical understanding of these objects is on firmer ground. If the warm and cool FYPS are actually two distinct types of pulsator with overlapping properties (as is the case for $\beta$ Cepheid variables and Slowly Pulsating B stars, which exhibit $p-$modes and $g-$modes respectively), then perhaps a new name for the warmer FYPS is warranted. On the other hand, if the warm FYPS are simply a subtype of either the FYPS or $\alpha$ Cygni variables, then a different naming scheme is necessary. As we discuss above, differences in post-main sequence luminosity correspond to differences in initial mass, and therefore evolutionary trajectory. For this reason, we use ``FYPS'' to refer to pulsating stars in our sample that are brighter than $\lum = 5.0$, and ``$\alpha$ Cyg'' to refer to the warmer, lower luminosity clump of pulsating stars identified previously.

\subsection{How real are these features in the HR diagram?}\label{subsec:bootstrap}

We now wish to determine the statistical significance of the features that we observe in the left panel of Figure \ref{fig:kde_fpulse_bootstrap}, in particular, the high fraction of pulsators above $\lum=5.0$. To do this, we need to establish our null hypothesis: that there is no underlying structure to the value of $f_{\rm pulse}$ as a function of luminosity and temperature, and that the patterns that we observe are simply a result of randomly labeling 36 stars from our sample as pulsators, as would be the case if contamination were responsible for our results (i.e., we would expect each lightcurve to have an approximately identical chance of being contaminated). If the null hypothesis were true, we could repeat the procedure described above, randomly taking 36 stars from the sample of 126, computing the KDE of these stars, and computing Eq. \eqref{eq:fpulse} using this simulated sample. We perform this experiment 10,000 times, and at each location in the HR diagram where we calculated $f_{\rm pulse}$, we compute the fraction of simulations in which the simulated value of $f_{\rm pulse}$ in that pixel is less than the observed value, a quantity we denote $f_{<f_{\rm pulse}}$. 

The top-right panel of Figure \ref{fig:kde_fpulse_bootstrap} shows $f_{<f_{\rm pulse}}$ as a function of position in the HR diagram. Pixels with high values of $f_{<f_{\rm pulse}}$ shown in deep red are temperatures and luminosities where the overdensity of pulsators are statistically unlikely to be a result of random sampling. Similarly, low values of $f_{<f_{\rm pulse}}$ shown in deep blue correspond to statistically significant {\it under}densities of pulsators. The important takeaway in this figure is that the patterns we discuss above are associated with extreme values of $f_{<f_{\rm pulse}}$; there is real structure in the value of $f_{\rm pulse}$ in the HR diagram. Furthermore, we find that the transitions from regions containing statistically more pulsators to regions containing fewer occurs fairly rapidly; the regions in orange and white that correspond to moderate values of $f_{<f_{\rm pulse}}$ are narrow. Finally, the overdensity of pulsators above $\lum=5.0$ is statistically significant (deep red) in nearly all locations in the plot.

Finally, we compute $f_{<f_{\rm pulse}}$ once more, limiting the calculation to stars in the LMC/SMC, which we show in the middle-right/bottom-right panels of Figure \ref{fig:kde_fpulse_bootstrap}. Similarly to the $f_{\rm pulse}$ calculation, while the exact structures are dependent on how well the HR diagram is sampled, the overall morphology and key features of the plot exist irrespective of the host galaxy. This is especially interesting for the SMC (bottom panels), where our sample is small, the \tess~lightcurves are significantly shorter, and targets are fainter on average. This indicates the robustness of both our results, and the procedure we use to identify pulsators. Unfortunately, as we conclude above, with such few stars in the SMC above $\lum=5.0$, we are unable to determine if there are any metallicity effects on the distribution of FYPS in the HR diagram.

\subsection{What about unaccounted for contaminants?}

The astute reader might ask: if so many of the frequencies recovered from the \tess~data are attributable to contamination from nearby stars, and the OGLE catalog is likely to not be 100\% complete, how can we conclude that contamination from nearby stars {\it not} listed in the OGLE catalog isn't the underlying cause of every single frequency that we measure? After all, the pulsations we are observing are incredibly fast relative to the dynamical timescale in a typical YSG, and the \tess~data are quite crowded. Are FYPS even real? We certainly believe so, and are convinced by the statistical significance of the overdensity of pulsators above $\lum=5$. 

Of course, more massive stars are found in more crowded regions \citep[e.g.][]{aadland18}, so one might argue that frequencies found in these more luminous stars are even more likely to be attributable to contamination. While we have no way of evaluating Eq. \eqref{eq:pfalse} without measured frequencies from nearby stars, we can look at the distribution of contaminant objects in the HR diagram. While the distribution of pulsators shown in Figure \ref{fig:kde_fpulse_bootstrap} does not appear to be drawn randomly from the overall sample, if it is similar to the distribution of contaminants, one can conclude that these pulsators are likely to be entirely an artifact of contamination. 

\begin{figure*}[ht!]
\includegraphics[width=\textwidth]{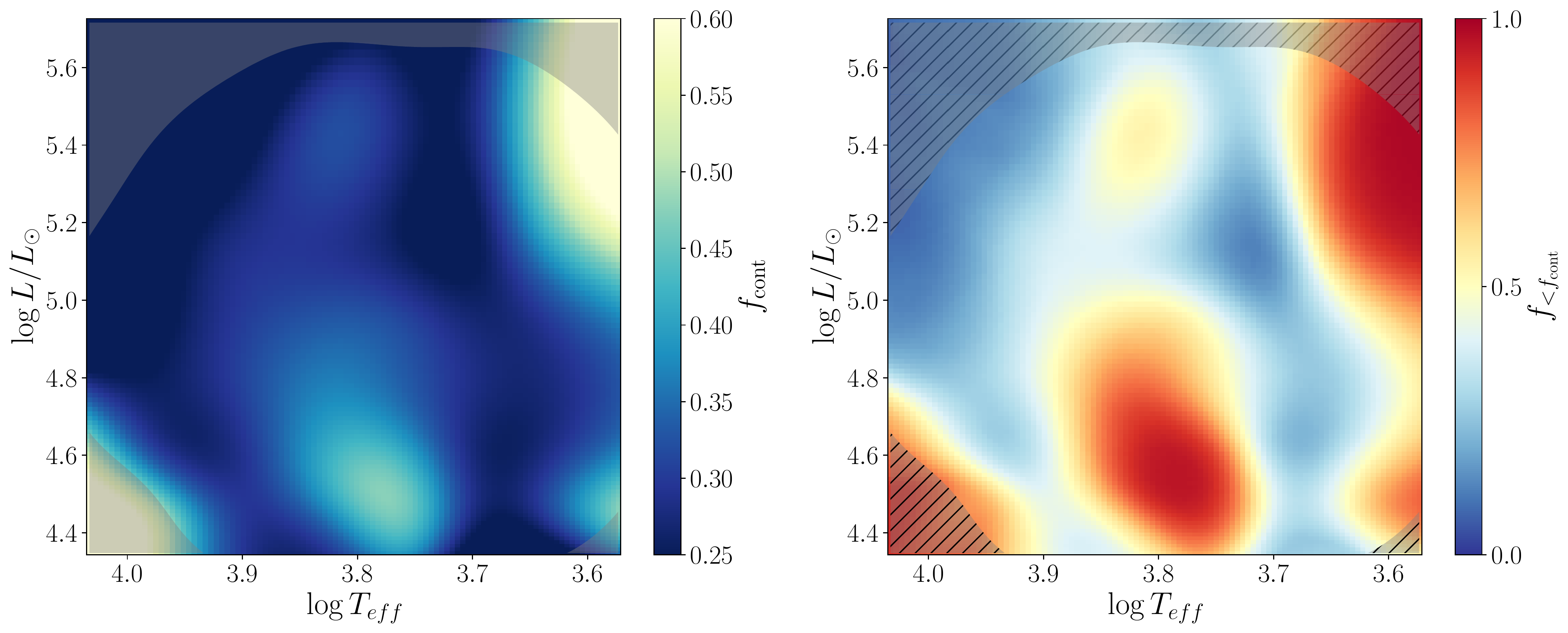}
\caption{Similar to Figure \ref{fig:kde_fpulse_bootstrap}, showing the distribution of contaminant objects in our sample. {\it Left panel}: The contaminant fraction $f_{\rm cont}$ shown as a function of location in the HR diagram, with $f_{\rm cont}$ increasing from blue to yellow. {\it Right panel}: An estimate of the statistical significance of the structures seen in the left panel. The fraction of simulated $f_{\rm cont}$ values drawn from 10,000 bootstrap simulations whose value is less than the true value of $f_{\rm cont}$, $f_{< f_{\rm cont}}$, with $f_{< f_{\rm cont}}$ increasing from red to blue. Regions in dark blue/red are statistically significant under/overdensities of contaminant objects.} \label{fig:kde_cont_bootstrap}
\end{figure*}

We begin by identifying the sample of objects for which every statistically significant frequency extracted from the \tess~data has a high likelihood of being a contaminant (i.e., the objects removed in \S\ref{subsubsec:ogle}). We can then use the KDE/bootstrap procedure described above and a variation of Eq. \eqref{eq:fpulse} to define the ``contaminant fraction'', $f_{\rm cont}$, as well as $f_{<f_{\rm cont}}$ by analogy with $f_{<f_{\rm pulse}}$. Figure \ref{fig:kde_cont_bootstrap} shows the results of these computations. With the exception of a small patch of contaminants at low luminosity and intermediate temperature, and one at high luminosity and low temperature, we find $f_{\rm cont}\approx=0.3$ and $f_{<f_{\rm cont}}\approx0.5$ throughout the HR diagram. In other words, in contrast with FYPS, the distribution of contaminant objects is statistically consistent with being drawn randomly from our overall sample of 126 stars. We see no evidence for an overdensity of contaminant objects above $\lum=5$, and conclude that it is unlikely that FYPS can be explained by unaccounted-for contamination. 

Of course, it is critically important to note the caveat that statistical tests like this are only useful to a point, especially in the small-number regime. Poorly-understood sources of contamination (namely, contamination from stars well-beyond the extent of a ``typical'' aperture used to extract \tess~lightcurves) remain a pitfall in the analysis of \tess~data of stars in crowded regions, and future observations will be crucial in order to confirm our conclusions.

\section{Discussion}\label{sec:discussion}

\subsection{Evolutionary Status: FYPS as Post Red Supergiants and the Progenitors of Type IIb Supernovae}\label{subsec:evostatus}

As we discuss in \S\ref{sec:intro}, massive stars that are first crossing the HR diagram are not expected to pulsate. We should only see pulsations excited in this part of the HR diagram in objects that (in a single-star paradigm) have experienced a prior RSG phase, and are now evolving leftward in the HR diagram as post-RSG objects. With a new understanding of the distribution of FYPS in the HR diagram, we can now ask the question: are FYPS post-RSGs? 

Two simple possibilities to identify them as such present themselves. First, we can search for signs of past or ongoing mass loss, either via a near-infrared excess from free-free emission in the wind, or a mid-infrared excess from circumstellar dust. Both of these are observed in the most luminous YSGs in M31 and M33 by \citet{gordon16}. However, these methods rely on the circumstellar material being detectable, which might not be the case for the lower luminosity stars in our sample. Indeed, we accessed 1.25-22 $\mu$m photometry of our targets from 2MASS \citep{cutri03}, the {\it Spitzer} SAGE survey \citep{meixner06}, and the WISE mission \citep{cutri13}. While there are individual high-luminosity FYPS that showed signs of a near- or mid-infrared excess, there are no systematic differences between FYPS and the non-pulsating YSGs above $\lum=5.0$. We are currently planning spectroscopic observations that will be more sensitive to ongoing mass loss and circumstellar material. The second possibility is to look for enhancements in the surface abundances of CNO-cycle products that may have been dredged up to the surface in a prior RSG phase, but detecting such enhancements will also require additional spectroscopy.

While we are currently obtaining these observations, we can examine the overall properties of FYPS for a hint as to their evolutionary status. The first clue is that $f_{\rm pulse}<1$ throughout the upper HR diagram; not all cool and luminous massive stars are FYPS. We use the binomial confidence interval \citep[][]{wilson27}, and find that $38.9 \pm {7.3}$ \% of stars in our sample brighter than $\lum\geq4.95$ are FYPS ($35.5 \pm 7.8$ \%, adopting $\lum\geq5.0$); by host galaxy, $41.4 \pm {8.3}$/$31.8 \pm {13.9}$ \% of high luminosity LMC/SMC stars are FYPS, a statistically insignificant difference. If FYPS are pre-RSG objects, then what separates FYPS from the non-pulsating YSGs with otherwise-identical surface properties that are presumably in the same evolutionary stage? Why don't {\it all} YSGs exhibit pulsations? On the other hand, if FYPS are post-RSGs, then they would have drastically different interior structures than the pre-RSG objects, potentially explaining why they pulsate.


We therefore posit that FYPS are indeed genuine post-RSG objects. Their minimum luminosity ($\lum \approx 5$) is then directly indicative of the minimum mass star that is capable of shedding its envelope during a prior RSG phase. It is also indicative of the {\it maximum} mass star that experiences core collapse as a RSG, resulting in a type II-P supernova explosion. In recent years, some debate has arisen in the literature regarding this mass threshold, $M_{h}$, with various statistical treatments of both the observed population of SN II-P progenitors and the theoretical landscape of explodability yielding values of $M_h$ between $\sim$17 and 25 $M_\odot$ \citep[e.g.][]{smartt09,sukhbold16,davies18,kochanek20}. If FYPS are post-RSGs, then the mass inferred from their minimum luminosity {\it should} correspond with $M_h$. However, if the inferred mass is too low (as in the case of the low luminosity $\alpha$ Cygni variables) or too high, then their status as post-RSGs would be ruled out.

In the {\it upper-}right panel of Figure \ref{fig:kde_fpulse_bootstrap}, we plot the 12, 15, 20, and 25 $M_\odot$ Geneva models at LMC metallicity \citep[$Z=0.006$;][]{eggenberger21} as solid black lines with their initial masses indicated. If post-RSGs evolve back across the HR diagram at the same luminosity as their pre-RSG counterparts, then the minimum FYPS luminosity corresponds to an initial mass of $\sim$16 $M_\odot$. However, a star's luminosity increases during the RSG phase. Depending on the exact mass loss mechanism that causes massive RSGs to shed their envelopes, the luminosity of a post-RSG might be somewhat higher. Therefore, we also use the end-of-life mass-luminosity relation presented by \citet{kochanek20} to plot the maximum luminosity $L_{\rm max}$ obtained by a 17 $M_\odot$ and a 21.3 $M_\odot$ star as dark and light blue lines respectively. The former value is consistent with the lower estimates of $M_h$ found in the literature \citep[e.g.][]{smartt09}, while the latter value is the Bayesian estimate of $M_h$ taken directly from Table 1 of \citet{kochanek20}; as noted by that author, the derived value of $M_h$ is biased high by some 3.3 $M_\odot$, implying that the true value of $M_h$ is closer to 18 $M_\odot$. 

\begin{figure}[t!]
\plotone{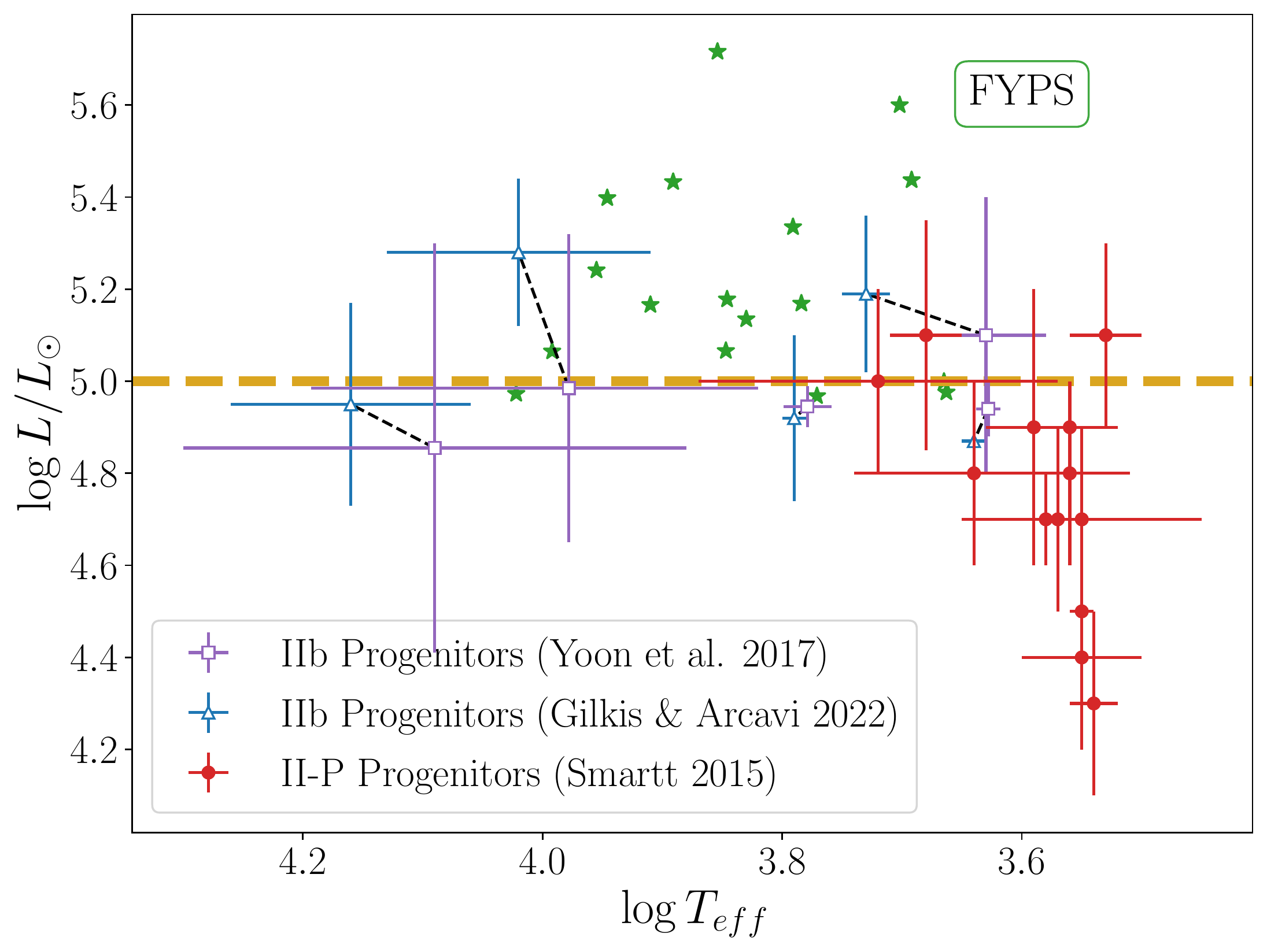}
\caption{Progenitor properties of 13 SNe II-P/II-L compiled by \citet{smartt15} in filled red circles, and five SNe IIb, with properties inferred by pre-explosion imaging and compiled by \citet{yoon17} (and citations therein) in open purple squares, and properties from detailed binary stellar evolution modelling from \citet{gilkis22} in open blue triangles. Black dashed lines link the properties of the progenitors of individual SNe inferred by each method. The stars we identify as FYPS are plotted as green stars, with the minimum FYPS luminosity of $\lum = 5.0$ shown as a dashed goldenrod line. The properties of supernovae progenitors and the transition from SNe II-P/II-L to SNe IIb compare favorably with the minimum luminosity of FYPS.} \label{fig:IIb_progenitors}
\end{figure}

Collectively, the evolutionary tracks and end-of-life luminosities serve to bracket the possible masses that a luminosity threshold at $\lum=5$ could correspond to. In any case, the minimum luminosity FYPS is consistent with $M_h\lesssim20 M_\odot$, especially considering that the luminosity of post-RSG stars above $M_h$ likely falls between their pre-RSG luminosity and $L_{\rm max}$. We therefore present the following evolutionary scenario for stars with masses between $M_h$ and $\sim$25 $M_\odot$: after leaving the main sequence, the star crosses the HR diagram and is observed as a RSG whose luminosity steadily rises. At some point during the RSG phase, some enhancement of the mass loss rate occurs (either through steady-state winds, episodic processes, or mass transfer onto a binary companion) that causes the star to lose enough of its envelope that it begins evolving leftward across the HR diagram at constant luminosity. Stars in this evolutionary stage are observed as FYPS. 

To begin to prove this picture where all stars above $\lum=5$ become post-RSGs, we can conduct one simple experiment by thinking about what happens when such a post-RSG undergoes core collapse. Stars that end their lives as RSGs produce type II-P or II-L supernovae (i.e., supernovae with strong H lines in their spectra), while partial or complete envelope stripping results in SNe IIb/Ib/Ic. Detailed radiative transfer modelling of SN IIb has inferred low ejecta masses and H/He mass fractions in these partially-stripped supernovae  \citep[e.g.][]{dessart11,yoon10}, implying that their progenitors cannot be much more massive than the progenitors of supernovae II-P \citep{smith14}.\footnote{We note that while evolutionary modelling with standard RSG mass loss rates predicts that all stars should lose multiple solar masses of material before core collapse, complete envelope stripping is not predicted at such low masses. As a result, early or late case B mass transfer in a binary system has been invoked to explain the properties of SN IIb progenitors \citep{yoon17}; however, if luminous RSGs truly do lose mass at significantly higher rates \citep{humphreys20}, binary interactions may not be necessary.} 

If FYPS are partially-stripped\footnote{regardless of whether or not this stripping occurs via single- or binary-star channels} cool supergiants with comparable masses to the most massive II-P supernova progenitors, then it becomes tempting to assert that FYPS are the progenitors of SNe IIb. If so, we would expect to observe supernovae properties transition from SNe II-P/II-L to SNe IIb for progenitors at the minimum FYPS luminosity. Furthermore, due to the initial mass function, and the fact that more massive stars likely experience greater degrees of envelope stripping, we would expect to see the properties of SNe IIb progenitors detected in pre-explosion imaging to cluster around the minimum FYPS luminosity. 

Figure \ref{fig:IIb_progenitors} shows the locations in the HR diagram of 13 SNe II-P and II-L as well as five SNe IIb progenitors with pre-explosion imaging. The red points show SNe II-P/II-L progenitors properties compiled by \citet{smartt15}. Orange points show the progenitor properties of SNe IIb inferred from the pre-explosion imaging and compiled by \citet{yoon17} (and references therein), while the blue points show the progenitor properties of the same sample of IIb supernovae inferred by recent detailed binary stellar evolution computations performed by \citet{gilkis22}. For comparison, we show both the minimum FYPS luminosity $\lum = 5.0$ as a dashed goldenrod line, and the stars we identify as FYPS as green stars. As we expect, the transition from SN II-P/II-L to SNe IIb occurs at the minimum luminosity boundary of FYPS, with the progenitors of IIb supernovae lined up neatly along this luminosity boundary, indicating that the scenario we presented above is viable. This finding is consistent with the evolutionary models presented by \citet{groh13}, who find that SNe II-L and IIb come from a roughly even mixture of yellow supergiant and luminous blue variable (LBV) progenitors with initial masses between 17 and 25 $M_\odot$. These classifications of their models were based on radiative transfer simulations; future spectroscopic observations will reveal whether any FYPS display features consistent with LBVs. 

We do caution against overinterpretation of this figure, especially the clustering of their luminosities around the minimum FYPS luminosity. Depending on the methodology used and conversion from pre-SN luminosity to initial mass, the initial masses of the SN IIb progenitors span a range between roughly 18 and 22 $M\odot$ (with sizeable uncertainty both from the luminosity estimates and the conversion from pre-SN luminosity to initial mass). If SNe IIb represent a population of partially-stripped stars with initial masses between 18 and 25 $M_\odot$ that might be observed as FYPS, then roughly 66\% of SN IIb progenitors should fall within the range of masses/luminosities already observed, assuming a standard \citet{salpeter55} IMF: with five SN IIb progenitors between 18 and 22 $M\odot$, we would expect between 2 and 3 SN IIb higher mass progenitors. The Poissonian probability of finding less than 1 such progenitor is $\sim$10\%; i.e., while relatively low probability, the lack of SN IIb progenitors spanning the entire mass range of FYPS is not statistically significant. If the clustering of SN IIb progenitors around $\lum\approx5.0$ does prove to be a real effect, this could be related to the fact that SNe IIb are produced by progenitors with just the right H envelope masses: too much produces a SN II-P/L, too little, a SN Ibc. As a result, we speculate that any real clustering of SN IIb progenitor luminosities may be reflective of the specific set of circumstances required to produce such progenitors.

An additional consideration is that while the majority of stars we identify as FYPS have moderate luminosities below $\lum=5.5$, we do identify a few pulsating stars with high luminosities that begin to approach the empirical upper-luminosity boundary observed in cool supergiants \citep{humphreys79}. This region of the HR diagram is also home to the ``yellow void,'' where the atmospheres of luminous supergiants become dynamically unstable \citep[e.g.][]{nieuwenhuijzen95}. One candidate FYPS in particular, HD 33579, is currently in the yellow void. Past work has shown that while it shares some similarities with other candidate post-RSGs, it (and notably-similar stars HD 7583 in the SMC and B324 in M33) are likely still on a redward evolutionary trajectory \citep{humphreys91,nieuwenhuijzen00,humphreys13,kourniotis22}. Therefore, the pulsations we observe in HD 33579 may be attributed to its dynamically unstable atmosphere

We also note that a few SNe Ibc now have claimed progenitor detections: PTF13bvn (Ib, discussed above), as well as SN2019yvr \citep[Ib;][]{kilpatrick21,sun22}, and SN2017ein \citep[Ib;][]{vandyk18}. One SN Ibc, SN2013ge, even has a claimed companion star for the progenitor \citep{fox22}. While in the case of SN2019yvr, \citet{kilpatrick21} place the progenitor in the region of Figure \ref{fig:IIb_progenitors} containing FYPS, \citet{sun22} perform detailed binary stellar evolution modeling and claim that the progenitor is a much lower luminosity object with a yellow hypergiant companion. Ultimately, more work (and more progenitors) are needed to understand how SN Ibc fit into the picture that we have outlined above, and so we refrain from including them in Figure \ref{fig:IIb_progenitors}.

Finally, we note that the discussion thus far has been mostly limited to a single-star perspective, whereas massive stars are preferentially born into binary systems \citep{sana12,sana13,moe17}. While Figure \ref{fig:kde_fpulse_bootstrap} reveals locations in the HR diagram that contain more pulsators on average, there is no location where we find zero pulsators. We can interpret this finding in the context of binary stellar evolution: at least some stars at all luminosities/initial masses covered by our sample lose their envelopes through binary interactions (at which point we can observe them pulsating),\footnote{Indeed, it is otherwise difficult to explain why most of the stars in the periwinkle ellipse in the top-left panel of Figure \ref{fig:kde_fpulse_bootstrap} are pulsators, as there are quite unlikely to be that many post-RSG objects at such a low luminosity. We speculate that these objects may be the products of binary interactions, but more observational evidence is needed to make such a conclusion.} while {\it all} stars with post-main sequence luminosities above $\lum\approx5$ lose their envelopes either through stellar winds or binary interactions. 

\subsection{Potential for Asteroseismic Studies}\label{subsec:astero}

The structure and evolution of massive stars is strongly dependent on the assumed physics \citep{martins13,farrell21}. A golden age of asteroseismology ushered in by space-based missions like {\it Kepler} and \tess~has revolutionized our understanding of these physics in main sequence massive stars --- from the interior mixing profiles in both the near-core region and the envelope \citep{michielsen19,pedersen21} to hints at angular momentum transport \citep{aerts19}. The discovery and characterization of FYPS represents the exciting possibility of probing the interiors of massive stars well after they have left the main sequence. However, this possibility remains beyond our reach at the present due to the fact that we have no understanding of what drives these pulsations, what regions of the star they might probe, or even whether they are $p-$modes, $g-$modes, or something else entirely.

\begin{figure}[t!]
\plotone{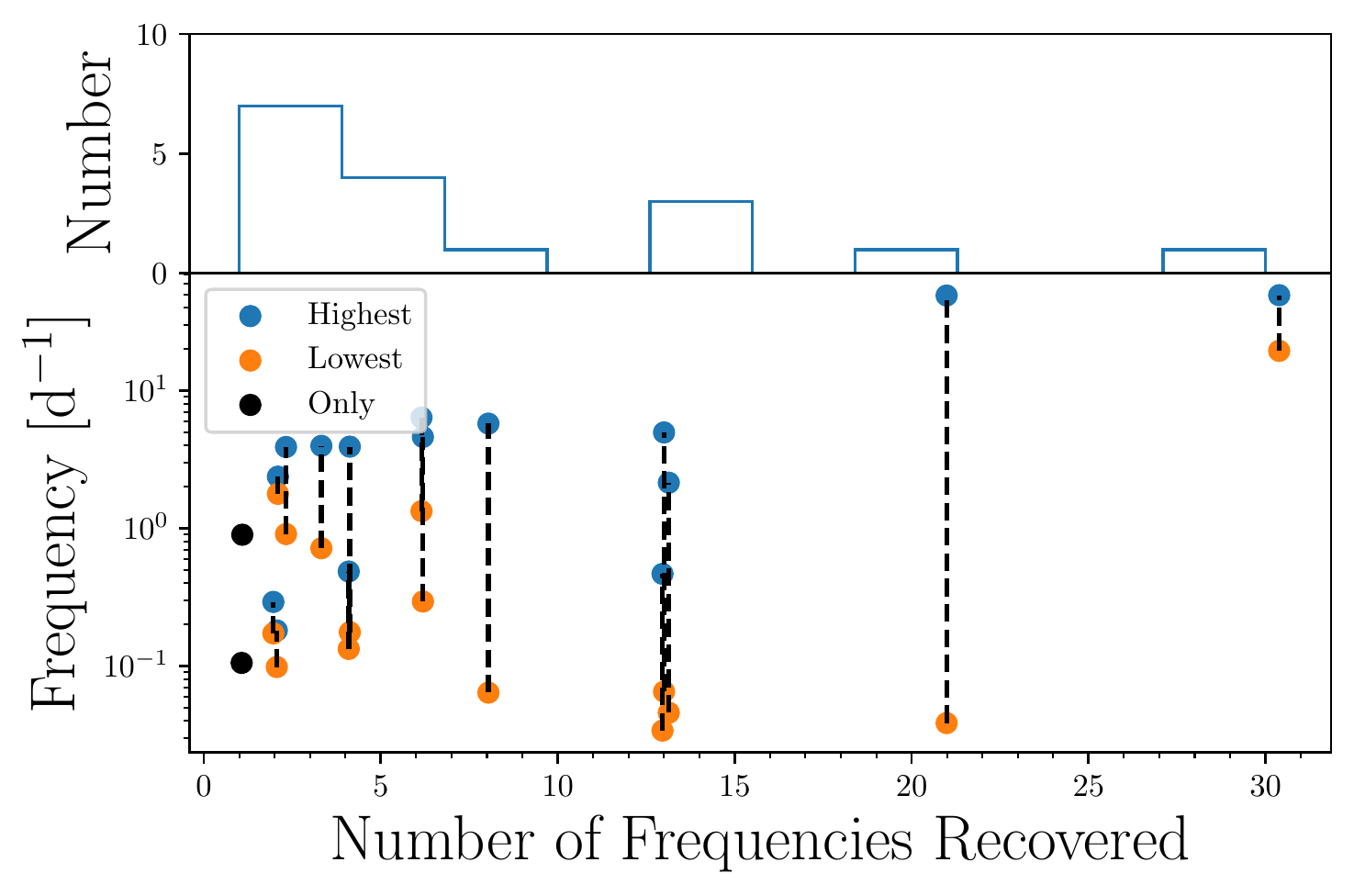}
\caption{{\it Top}: Histogram showing the number of FYPS from which a given number of frequencies have been extracted. {\it Bottom}: Number of frequencies extracted via prewhitening versus the highest (blue) and lowest (orange) extracted frequency, with lines connecting points belonging to the same star. Stars with only one frequency are shown as black points. The majority of FYPS have 10 or fewer frequencies spanning approximately 0.1-10 d$^{-1}$.} \label{fig:freq_props}
\end{figure}

A detailed theoretical treatment of pulsations in FYPS is ongoing (and beyond the scope of this work). However, we can take an early look at the behavior of the modes we observe in FYPS to glean a hint of what we are seeing. First we examine the distribution of pulsation properties. The top panel of Figure \ref{fig:freq_props} shows the number of stars from which a given number of frequencies have been extracted, while the bottom panel shows the highest and lowest extracted frequency in blue/orange respectively, versus the number of frequencies extracted; points belonging to the same star have been connected by a dashed black vertical line. For stars with only one frequency, we show only that frequency in black. For clarity, a logarithmic scaling has been applied to the y-axis, and a small random horizontal offset has been applied to each set of points. The majority of FYPS in our sample all display similar properties, with 10 or fewer recovered frequencies that fall in the approximate range between 0.1 and 10 d$^{-1}$. Two stars, HV 829 and HD 269661, display notably different behavior, with over 20 frequencies that extend well beyond 10 d$^{-1}$. Given the difference between these objects and the other FYPS, we ignore them for the remainder of the analysis.

\begin{figure*}[ht!]
\plotone{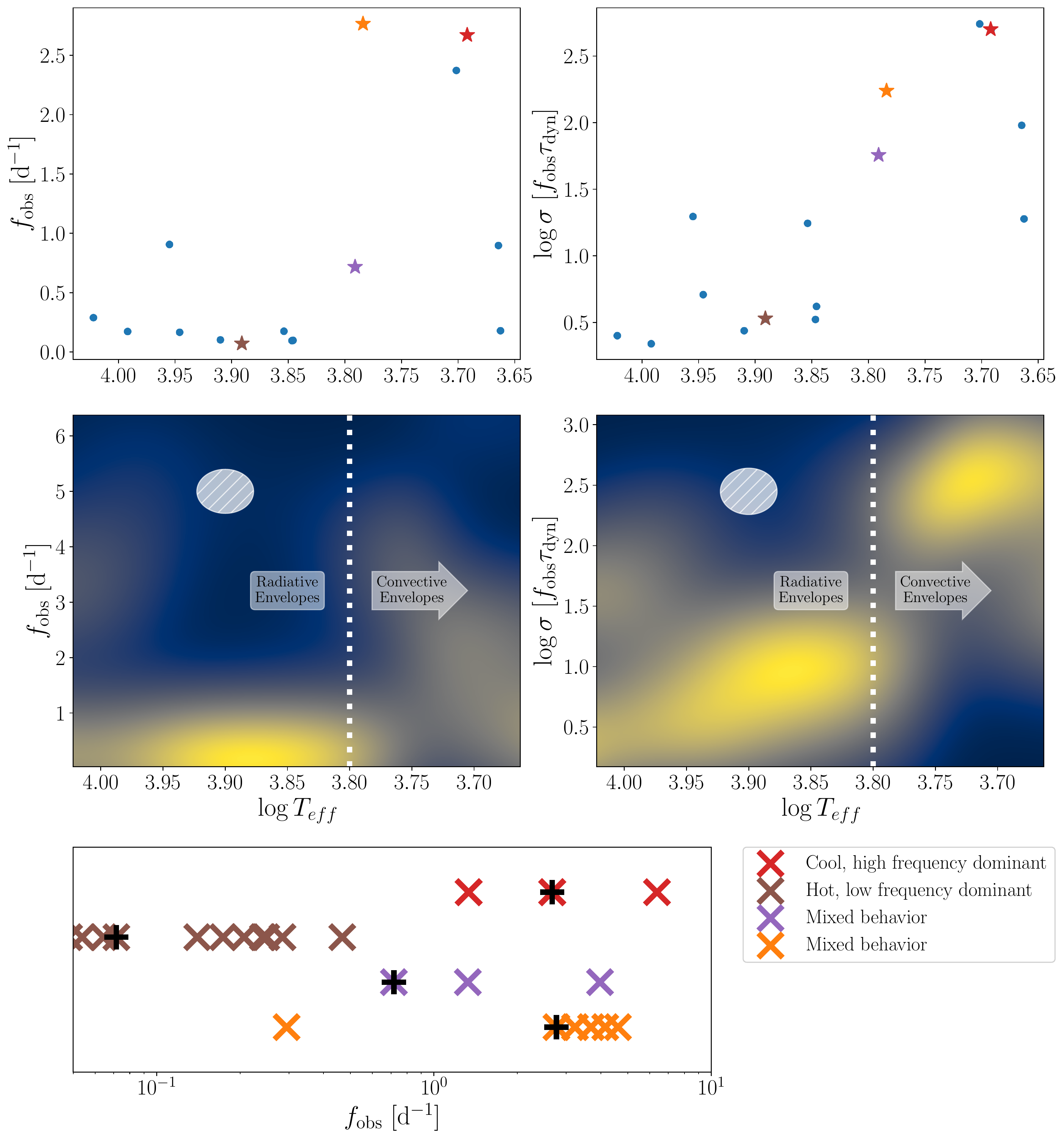}
\caption{{\it Upper Left}: Scatter plot showing the dominant frequency in each FYPS as a function of effective temperature, after filtering out stars with more than ten frequencies or with any frequencies above 10 d$^{-1}$. {\it Upper Right}: Similar, but for dimensionless frequencies calculated by multiplying the observed frequencies for each star by an estimated dynamical timescale based on the star's position in the HR diagram, assuming a mass of 15 $M_\odot$. {\it Center Left}: A KDE showing the distribution of observed FYPS frequencies for the same sample as a function of effective temperature. The KDE has been scaled by the distribution of FYPS in temperature-space to highlight the distribution of FYPS frequencies. {\it Center Right}: Similar, but for the dimensionless frequencies. For both lower panels, we show the shape of the KDE kernel in the space as a white ellipse, and a vertical line at $\mathbf{\teff=3.8}$ showing where H$^{-}$ opacity begins to contribute significantly to the opacity of stellar atmospheres. To the left of this line, stellar envelopes are radiative, while to the right, stars begin to develop increasingly deep convective envelopes. This physical transition corresponds to a difference in the observed properties of FYPS pulsations. {\it Bottom}: Frequencies recovered by iterative prewhitening of four example FYPS. Red points show a cool FYPS with high pulsational frequencies, and brown points show a warm FYPS with low pulsational frequencies. Two FYPS at $\teff\approx3.8$ are shown, both with frequencies spanning the range of those seen in FYPS, but one with a low dominant frequency (purple), and one with a high dominant frequency (orange). The y-coordinate is an arbitrary value to separate each star's frequencies, and black pluses indicate the dominant frequency in each star. We highlight the locations of these stars in the upper panels with correspondingly-colored star markers.} \label{fig:freq_scaling}
\end{figure*}

The upper left panel of Figure \ref{fig:freq_scaling} shows the dominant frequency recovered in this sample of FYPS as a function of effective temperature. Two distinct behaviors can be seen: one group of FYPS where the dominant frequency is steeply correlated with temperature, and one in which the dominant frequency is much lower ($<0.5$ d$^{-1}$). Another way of visualizing these data is by instead plotting dimensionless frequencies $\sigma$, defined as:
\begin{equation}
   \sigma = f_{\rm obs}\tau_{\rm dyn} \approx \frac{f_{\rm obs}}{\sqrt{G \bar{\rho}}}
\end{equation}
where to estimate the average density, $\bar{\rho}$, we derive the radius of each star from its position in the HR diagram, and assume a fiducial mass of 15 $M_\odot$ for all stars. We note that this assumption does not introduce an inordinate amount of uncertainty, due to the fact that $\tau_{\rm dyn}$ depends most strongly on effective temperature via the radius ($\tau_{\rm dyn}\propto R^{3/2} \propto T_{\rm eff}^{-3}$). Furthermore, the mass of each star is likely within a factor of two of the fiducial mass, whereas the inferred radii of stars in our sample varies by almost a factor of ten across the temperature range of our sample. To highlight the large dynamic range of inferred $\sigma$ values, we apply a logarithmic scaling to the y-axis in this panel. Once again, two behaviors can be seen, with one group displaying high dominant frequencies, and the second group displaying low dominant frequencies. The transition between the regimes seen in these panels occurs at $\teff=3.8$. 


Of course, these are only the dominant frequency recovered from each star. Perhaps more interesting is the overall {\it distribution} of recovered frequencies. To visualize this, we adapt the procedure we used to compute $f_{\rm pulse}$ in \S\ref{sec:results} to derive the distribution of frequencies as a function of effective temperature. We first transform the observed variables ($\teff$ and observed frequency) to have zero mean and unit variance, before computing the kernel density estimate. However, this KDE is also reflective of the distribution of FYPS in $\teff$, so we also compute a one-dimensional KDE of the FYPS effective temperatures, and divide each row of the two-dimensional KDE by this one-dimensional KDE. Both KDEs have Gaussian kernels with a bandwidth of 0.5 in the transformed variables. The center left panel of Figure \ref{fig:freq_scaling} shows the result of this calculation; regions in yellow contain more observed frequencies, while regions in blue contain fewer. The white ellipse shows the shape of the Gaussian kernel transformed into the observed variables. 

Again, we recover two distinct behaviors. At high temperatures, FYPS predominantly pulsate at lower frequencies, while at cooler temperatures, we find a diagonal cloud of frequencies that is steeply correlated with effective temperature. The transition between these regimes again occurs around $\teff=3.8$. We note that this is the exact region of the HR diagram where $f_{\rm pulse}$ decreases in the top-left panel of Figure \ref{fig:kde_fpulse_bootstrap}. This temperature range corresponds to where stars begin to develop convective envelopes; in main sequence stars, H$^{-}$ opacity becomes the dominant opacity source around 8000 K, and a strong convective envelope is strongly developed around 6,000 K \citep{vansaders12}. We indicate the latter temperature as a dashed white vertical line, and denote the regions of the plot in which stars have mostly radiative or convective envelopes. Extending between these two regions is the diagonal structure that we also see in the upper left panel of the figure, and the transition across the two observed behaviors is a smooth one. The center right panel of Figure \ref{fig:freq_scaling} shows the distribution of frequencies in $\sigma$ versus $\teff$. We once again see a region of low dimensionless frequencies at high temperature, and a smooth transition to higher dimensionless frequencies.

One interpretation of Figure \ref{fig:freq_scaling} is that we are perhaps seeing two types of pulsating stars occupying the same part of the HR diagram: one group with low frequency pulsations that are seen at temperatures above $\teff\gtrsim3.8$, and another with higher frequency pulsations that span the entire temperature range of the sample. Indeed, we can find examples of FYPS where this behavior is born out; the red and brown X's in the bottom panel of Figure \ref{fig:freq_scaling} show the frequencies recovered from two FYPS, with an arbitrary vertical offset applied to allow for each star's frequencies to be seen clearly. The red example is a cool FYPS; typical for FYPS in this part of the HR diagram, we recover three individual frequencies between one and ten cycles per day. The brown example is a FYPS on the lower ``branch'' of hot FYPS, and as expected, it only shows low frequencies below 1 d$^{-1}$.

However, this picture is complicated by the other two example FYPS that we show. The purple points in the bottom panel are the frequencies from a star on the ``lower'' branch of hot FYPS, and the frequencies recovered from a hot FYPS on the ``upper'' branch are plotted in orange. While each star's dominant frequency belongs to a different branch from the upper-left panel of Figure \ref{fig:freq_scaling}, both stars display both high and low frequencies; the high frequencies are consistent with the frequencies observed in the cool FYPS (shown in red). For reference, we show the location of all four example FYPS in the upper panels of Figure \ref{fig:freq_scaling} as correspondingly-colored star-shaped points.

Another interpretation of Figure \ref{fig:freq_scaling} is based on the fact that the transition between the two observed behaviors occurs at roughly the same effective temperature at which stars begin to develop convective outer envelopes. Because $g-$modes cannot propagate through convective layers, we propose that the lower-temperature FYPS display higher-frequency $p-$modes that propagate through the outer layers of the star, including the convective envelope. In this scenario, as the star loses progressively more mass and evolves from cooler to hotter temperatures (right to left in this plot), the convective envelope shrinks until the outer layers of the star become radiative, and lower-frequency $g-$modes that would otherwise be confined to the interior can be observed at the surface (in addition to higher-frequency $p-$modes). If this were the case, $p-$modes in the convective envelope of a FYPS might be able to couple with $g-$modes of similar frequency that are confined to the interior of the star; preliminary modelling work in \citet{dornwallenstein20b} showed that the typical observed frequencies could propagate as $g-$modes in the interior of a post-RSG model, and that the evanescent region between the $g-$ and $p-$ mode cavities is incredibly thin, lending credence to this scenario. Such mixed modes could possibly be used to probe the entire structure of the star from the outer boundary of the convective core to the surface. This prospect is incredibly exciting, and we encourage the community to invest significant effort in understanding FYPS.

\section{Conclusions}\label{sec:conclusion}

Our main results are summarized as follows:

\begin{itemize}
  \item From a sample of 126 cool supergiants with confirmed membership in the Magellanic Clouds, we identify 36 stars as pulsators after making quality cuts and removing contaminants. These pulsators reside in two regions in the HR diagram: one region at low luminosity and high temperature that we identify as $\alpha$ Cygni variables, and one region at luminosities above $\lum = 5.0$ that we associate with fast yellow pulsating supergiants. Using a bootstrap analysis, we find that these structures in the HR diagram are real and cannot be explained by random chance (i.e., unaccounted-for contamination by nearby stars or binary companions).
  \item $\lum = 5.0$ corresponds quite well with the maximum initial masses of SNe II-P progenitors detected through pre-explosion imaging. This, combined with the fact that the inferred properties of SNe IIb progenitors are well-matched with the properties of FYPS and the fact that only $\sim$40\% of stars above $\lum = 5.0$ pulsate leads us to conclude that FYPS are likely post-RSG objects. We are currently conducting a spectroscopic observational campaign to investigate this further.
  \item No models for FYPS pulsations currently exist. However, by examining the properties of FYPS pulsations, we present a scenario, wherein at least some subset of FYPS are mixed-mode pulsators, whose pulsations may be used to probe the entirety of the stellar structure from the surface down to the edge of the convective core. We are currently working on a theoretical study to ascertain the likelihood of this scenario. In any case, asteroseismic analyses of massive stars in such an evolved state would revolutionize our understanding of the late phases of massive stars evolution, and we strongly encourage the community to work towards understanding these fascinating objects.
\end{itemize}

\acknowledgments

The authors acknowledge that the work presented was largely conducted on the traditional land of the first people of Seattle, the Duwamish People past and present and honor with gratitude the land itself and the Duwamish Tribe. 

We gratefully acknowledge C. Kochanek for insightful discussions about contamination from nearby stars in the \tess~data.

This research  was  supported  by  NSF  grant AST 1714285 and a Cottrell Scholar Award from the Research Corporation for Scientific Advancement granted to EML. 

KFN acknowledges support from the Dunlap Institute.

This work has been carried out in the framework of the PlanetS National Centre of Competence in Research (NCCR) supported by the Swiss National Science Foundation (SNSF). 

JRAD and KAB acknowledge support from the DIRAC Institute in the Department of Astronomy at the University of Washington. The DIRAC Institute is supported through generous gifts from the Charles and Lisa Simonyi Fund for Arts and Sciences, and the Washington Research Foundation.

The authors gratefully acknowledges support from the Heising-Simons Foundation, and from the Research Corporation for Science Advancement for hosting the 2019 Scialog meeting on Time Domain Astrophysics.

This research has made use of the SIMBAD database, operated at CDS, Strasbourg, France. This research has made use of the VizieR catalogue access tool, CDS, Strasbourg, France (DOI: 10.26093/cds/vizier). The original description of the VizieR service was published in A\&AS 143, 23. 

This work has made use of data from the European Space Agency (ESA) mission
{\it Gaia} (\url{https://www.cosmos.esa.int/gaia}), processed by the {\it Gaia}
Data Processing and Analysis Consortium (DPAC,
\url{https://www.cosmos.esa.int/web/gaia/dpac/consortium}). Funding for the DPAC
has been provided by national institutions, in particular the institutions
participating in the {\it Gaia} Multilateral Agreement.

The \tess~data described here for Sectors 1-13 and 27-39 may be obtained from the MAST archive at
\dataset[doi:10.17909/t9-nmc8-f686]{10.17909/t9-nmc8-f686}.

This work made use of the following software and facilities:

\vspace{5mm}

\facility{The {\it Transiting Exoplanet Survey Satellite} (\tess, \citealt{ricker15})}

\software{
Astropy v4.3.1 \citep{astropy13,astropy18},
Astroquery v0.3.10 \citep{astroquery},
Celerite2 v0.2.0 \citep{foremanmackey17,foremanmackey18},
Matplotlib v3.5.1 \citep{Hunter:2007}, 
NumPy v1.21.5 \citep{numpy:2011}, 
Pandas v1.3.4 \citep{pandas:2010}, 
PyMC3 v3.11.4 \citep{salvatier16},
PySynphot v1.0.0 \citep{pysynphot13},
Python 3.7.4, 
Scikit-learn v1.0.2 \citep{scikit-learn11},
Scipy v1.7.3 \citep{scipy:2001,scipy:2020},
Theano v0.0.4 \citep{theano16}
}

\newpage
\bibliography{bib}
\bibliographystyle{aasjournal}

\appendix
\restartappendixnumbering

\section{Identifying Pulsating Stars With a Gaussian Process}\label{app:GP}

In \S\ref{subsec:identify}, we briefly describe the procedure we use to identify stars with periodic variability superimposed onto the stochastic low frequency background. Here we describe this procedure in full. We first perform the iterative prewhitening procedure that we developed in \citet{dornwallenstein20b}, which computes the periodogram of the \tess~lightcurve, fits the shape of the SLF variability using Eq. \eqref{eq:rednoise}, identify the highest peak in the periodogram, fit the lightcurve with a sinusoid at the frequency of the highest peak and the first two harmonics of the sinusoid, subtract the best-fit model from the data, and repeat, progressively adding sinusoids (and harmonics) onto the model for the lightcurve until a minimum in the Bayesian Information Criterion is reached. In addition to the best-fit frequencies, amplitudes, and phases, we also derive standard errors using the equations from \citet{lucy71} and \citet{montgomery99}. These errors also allow us to define a signal-to-noise ratio using the best-fit value of the amplitude and its associated error for a given frequency. The results of this prewhitening procedure (as well as all code needed to reproduce this work) are available online in HDF5 format at \url{https://github.com/tzdwi/TESS}.

Though we account for the shape of the SLF variability in the frequency domain, the fit to the lightcurve occurs in the time domain, where the SLF variability is still a significant portion of the overall variability. As a result, the identified frequencies may be spurious detections, especially for lightcurves with only one or two \tess~sectors (i.e., stars in the SMC). Even though a model with a sum of sinusoids might result in a lower BIC than fitting the data with a constant, this tells us nothing about whether a fit with a stochastic model might be more appropriate. Furthermore, the derived frequencies and errors might be biased by the unaccounted-for SLF variability.

We instead attempt to search for statistically significant freqencies while simultaneously marginalizing over a stochastic process. We use {\sc celerite2} \citep{foremanmackey17,foremanmackey18}, which implements GPs with covariance matrices whose entries can be calculated from a kernel function written as a sum of exponential terms. Following \citet{bowman22}, we use the {\tt SHOTerm} kernel, which models the variability as a stochastically-driven, damped harmonic oscillator, and has also been used to model granulation in sun-like stars \citep[e.g.][]{pereira19}. The power spectrum corresponding to this kernel is commonly written
\begin{equation}\label{eq:SHO_spec}
    S(\omega) = \sqrt{\frac{2}{\pi}}\frac{S_0\omega_0^4}{(\omega^2-\omega_0^2)^2 + \omega^2\omega_0^2/Q^2}
\end{equation}
with the angular frequency $\omega$, the undamped angular frequency of the oscillator $\omega_0$, a term to capture the amplitude of the variability $S_0$, and a ``quality factor'' $Q$. We instead adopt a parametrization with $\rho=2\pi/\omega_0$ being the undamped period, $\tau 2Q/\omega_0$ being the timescale on which perturbations are damped, and $\sigma=\sqrt{S_0\omega_0 Q}$ being the standard deviation of the variability. As demonstrated by \citeauthor{bowman22}, these parameters correspond well with $\tau_{\rm char}$, $\gamma$, and $\alpha_0$ from Eq. \eqref{eq:rednoise}, respectively. To the diagonal of the covariance matrix $K_{\pmb{\alpha}}$, we also add a vector containing the squared uncertainties of each point in the lightcurve, and a constant to account for any excess white noise in the data (a.k.a., jitter, corresponding to $\alpha_w$).

After applying the initial prewhitening procedure to identify stars with and without significant frequencies, we fit the periodograms with Equation \eqref{eq:rednoise} to obtain initial guesses for the SLF parameters. We then transform these initial guesses into the parametrization described above by equating Equation \eqref{eq:SHO_spec} with the square of Equation \eqref{eq:rednoise}, assuming the quality factor $Q = 1/\sqrt{2}$ as shown in the third column of Table \ref{tab:gp_priors}. For each parameter, we apply a uniform prior in log space, with boundaries indicated in the fourth column of Table \ref{tab:gp_priors}. We deliberately choose a very wide prior for each parameter. 

For the stars without significant frequencies, we first maximize the log-likelihood in Eq. \eqref{eq:GP_ll} using {\sc PyMC3} \citep{salvatier16}, which is compatible with {\sc celerite2}. For the mean function $\pmb{\mu_{\theta}}$, we fit a constant term; because the lightcurves have been normalized by the median, this constant is always very close to 1. After deriving the maximum-likelihood values for the SLF parameters, the mean flux, and the jitter, we then calculate the BIC as 
\begin{equation}
    BIC_{\rm SLF} = -2\log \mathcal{L} + 5\ln N
\end{equation}
where 5 is the number of parameters at this stage, and $N$ is the number of points in the lightcurve \citep{schwarz78}. We then compute the posterior prediction of the flux using {\sc celerite2}, and subtract the result from the lightcurve. To propagate our errors, we also compute the variance of the GP, and add the square root of the variance to the uncertainties on each data point in quadrature. Because computing the GP variance at each observation is quite computationally expensive, we instead compute at every 50$^{\rm th}$ data point (i.e., every 100 minutes), and linearly interpolate the result for the remaining data. In practice, the variance changes on timescales comparable with $\rho$ (of-order days), so this step results in only a minimal loss of information.

Next, we perform prewhitening on the residual flux (and associated uncertainties). If any frequencies are found in the residual light curve, we then iteratively fit the original light curve with the GP. However, rather than treating the mean flux as a constant, at the $i^{\rm th}$ stage in this process we maximize the log-likelihood in Eq. \eqref{eq:GP_ll}, with a mean model equal to a constant plus a sum of $i$ sinusoids whose frequencies, amplitudes, and phases are initialized from the first $i$ highest-amplitude frequencies recovered by prewhitening. We bound each frequency and amplitude to lie in a narrow range contained within $\pm3$ times the errorbar for each parameter returned by the prewhitening procedure (following \citealt{lucy71} and \citealt{montgomery99}; described in further detail in \citealt{dornwallenstein20b}), with the additional consideration that the amplitude must always be positive. The phase is only constrained to fall between $-\pi$ and $\pi$. We then calculate the BIC as
\begin{equation}
    BIC_{\rm i} = -2\ln \mathcal{L} + (5+3i)\ln N
\end{equation}
where now there are $3i$ parameters per frequency in the model, plus the original five for the SLF variability, jitter, and mean flux. If at any stage, $BIC_i < BIC_{\rm SLF}$, then the model with both SLF variability and coherent frequencies is preferred over the model with just SLF variability, and we identify the star as being periodically variable. 

For the stars {\it with} frequencies identified in the initial prewhitening run, we perform a similar procedure, computing $BIC_{\rm SLF}$ and then iteratively adding the recovered frequencies until $BIC_i < BIC_{\rm SLF}$. If $BIC_i \geq BIC_{\rm SLF}$ at all stages, we use the SLF variability-only model to compute the posterior prediction for the flux as well as the variance, subtract this prediction from the flux, run prewhitening on the residuals, and once again iteratively add any frequencies recovered to the mean model. In practice, we found that there were no stars without frequencies recovered by the initial prewhitening run from which we later recovered significant frequencies after subtracting off the SLF variability predicted by the GP. However, there was a single star initially identified as periodically variable for which the SLF variability-only model was preferred. Furthermore, we note again that the structures we see in the bottom panels of Figure \ref{fig:kde_fpulse_bootstrap} are qualitatively similar to the center and top panels, despite the SMC sample containing fewer (and generally fainter) stars, the \tess~lightcurves of these stars being significantly shorter, and the overall poor sampling of the HR diagram in the SMC. Finally, while testing this procedure, when we fit all ten frequencies initially identified in the lightcurve of the FYPS HD 269953, computed the posterior prediction of the flux (including all frequencies), calculated the residual flux after subtracting off the GP, and ran prewhitening once more, we recovered an additional five low-amplitude frequencies that were otherwise undetectable beneath the SLF background; adding these frequencies to the model further reduced the BIC. These two facts imply that the method that we present here is both more robust at identifying pulsating stars in the presence of SLF variability (especially for stars with shorter and/or noisier \tess~lightcurves as in the SMC), and has the potential to derive more reliable pulsation frequencies and associated uncertainties, a critical ingredient in asteroseismic analyses. In order to do the latter, we would need to fully sample the posterior probability distribution of the entire model, which is a quite computationally intensive task. As the goal of the present work is simply to identify FYPS and characterize their properties, we will return to this point in a future paper. 

\begin{deluxetable*}{llcc}
\tabletypesize{\small}
\tablecaption{Summary of initial guesses and priors used to perform the GP regression. $\mathcal{U}(a,b)$ refers to a uniform prior with lower and upper limits of $a$ and $b$, while $\mathcal{N}(\mu,s)$ refers to a normal distribution with mean $\mu$ and standard deviation $s$. We note that for $\sigma$, 
$\rho$, $\tau$, and the jitter, the prior is uniform in log-space. Finally, $\alpha_0$, $\tau_{\rm char}$, and $\alpha_w$ refer to the best-fit values of these parameters from Equation \eqref{eq:rednoise}.\label{tab:gp_priors}}
\tablehead{\colhead{Parameter} & \colhead{Description} & \colhead{Initial Guess} & \colhead{Prior}} 
\startdata
$\sigma$ & Standard deviation of the stochastic variability. & $\sigma_0 = \alpha_0 \tau_{\rm char}^{-1/2} (\pi / 4)^{1/4}$ & $\mathcal{U}((\ln \sigma_0) - 10,(\ln \sigma_0) + 10)$ \\ 
$\rho$ & Undamped characteristic period. & $\rho_0 = 2\pi\tau_{\rm char}$ & $\mathcal{U}(\ln 0.01 \rho_0, \ln 100 \rho_0)$ \\
$\tau$ & Damping timescale & $\tau_0 = \sqrt{2}\tau_{\rm char}$ & $\mathcal{U}(\ln 0.01 \tau_0, \ln 100 \tau_0)$ \\
Mean flux & The mean of the lightcurve. & 1 & $\mathcal{N}(1, {\rm std(flux))}$\tablenotemark{a} \\
Jitter & Excess uncorrelated noise in the data. & $\alpha_w$ & $\mathcal{U}((\ln \alpha_w) - 15, (\ln \alpha_w) + 15)$ \\
$f^{i}$ & $i^{\rm th}$ frequency found via prewhitening. & $f^{i}_0$\tablenotemark{b} & $\mathcal{U}(f^{i}_0 - 3 \epsilon(f^{i}_0), f^{i}_0 + 3 \epsilon(f^{i}_0)$\tablenotemark{c} \\
$A^{i}$ & Amplitude of $f^{i}$. & $A^{i}_0$ & $\mathcal{U}({\rm max}(A^{i}_0 - 3 \epsilon(A^{i}_0), 0), A^{i}_0 + 3 \epsilon(A^{i}_0)$) \\
$\phi^{i}$ & Phase of $f^{i}$. & $\phi^{i}_0$ & $\mathcal{U}(-\pi,\pi)$ \\
\enddata
\tablenotetext{a}{Here std(flux) refers to the standard deviation of the lightcurve.}
\tablenotetext{b}{$f^{i}_0$ is the best-fit frequency derived via iterative prewhitening.}
\tablenotetext{c}{$\epsilon(f^{i}_0)$ is the error on the corresponding frequency, using the formulae given in \citet{lucy71} and \citet{montgomery99}, which were previously listed in \citet{dornwallenstein20b}. Similar formulae are used to compute errors on the amplitudes and phases.}
\end{deluxetable*}

\clearpage
\onecolumngrid
\section{FYPS, low-luminosity pulsators, and non-pulsators in our sample}\label{app:FYPS}

Tables \ref{tab:lmc_fyps} and \ref{tab:smc_fyps} list the TIC numbers, common names, coordinates, \tess~magnitudes, positions in the HR diagram, number of frequencies recovered via prewhitening ($N_f$), highest-amplitude recovered frequency ($f_0$), and literature spectral types of the stars identified as fast yellow pulsating supergiants in the LMC and SMC respectively. We also tabulate the pulsating stars below $\lum=5.0$ in Tables \ref{tab:lmc_acyg} and \ref{tab:smc_acyg}, as well as the non-pulsating stars in our sample in Tables \ref{tab:lmc_np} and \ref{tab:smc_np}. We note that the spectral types and effective temperatures are generally in alignment. In most cases of mismatches between the temperature from \citeauthor{neugent10} and the literature spectral type (e.g., HD 269860 in Table \ref{tab:lmc_acyg}; $\teff=3.787$, spectral type A3Ia), the latter is frequently old enough (in this case from \citealt{stock76}), that we can't rule out the possibility of spectroscopic variability without further observations. Figures B\ref{fig:lmc_fyps} and B\ref{fig:smc_fyps} show the \tess~lightcurves and residual power spectra of the LMC and SMC FYPS respectively. The residual power spectra are computed by normalizing the Lomb-Scargle periodogram by the square of Equation \ref{eq:rednoise}, adopting the best-fit parameters following \citet{dornwallenstein20b}. Frequencies identified by prewhitening are shown as vertical lines, with frequencies associated with contaminants in red, and genuine frequencies in grey.

\begin{deluxetable*}{lccccccccc}
\tabletypesize{\scriptsize}
\tablecaption{Names, TIC numbers, coordinates, \tess~magnitudes, positions in the HR diagram, number of frequencies recovered via iterative prewhitening after removing contaminants ($N_f$), highest amplitude recovered frequency ($f_0$),
and literature spectral types of the FYPS identified in the LMC, ordered by effective temperature from coolest to warmest. Typical uncertainties in $\log T_{\rm {eff}}$ and $\log L/L_\odot$ are 0.015 dex and 0.10 dex respectively.\label{tab:lmc_fyps}}
\tablehead{\colhead{Common Name} & \colhead{TIC Number} & \colhead{R.A.} & \colhead{Dec} & \colhead{$T$} & \colhead{$\log T_{\rm{eff}}$} & \colhead{$\log L/L_\odot$} & \colhead{$N_f$} & \colhead{$f_0$} & \colhead{Lit. Spectral Type}\\
\colhead{} & \colhead{} & \colhead{[deg]} & \colhead{[deg]} & \colhead{[mag]} & \colhead{[K]} & \colhead{$L_\odot$} & \colhead{} & \colhead{d$^{-1}$} & \colhead{}} 
\startdata
HD 269953 & 404850274 & 85.05069622 & -69.66801469 & 9.267 & 3.692 & 5.437 & 6 & 2.671 & G0 0 \citep{keenan89} \\ 
HD 269723 & 425083216 & 83.10401996 & -67.69822538 & 7.702 & 3.702 & 5.600 & 2 & 2.374 & G4 0 \citep{keenan89} \\ 
HD 268687 & 29984014 & 72.73273606 & -69.43125133 & 10.465 & 3.784 & 5.169 & 6 & 2.765 & F6Ia \citep{ardeberg72} \\ 
HD 269840 & 277108449 & 84.04200662 & -68.92812902 & 10.132 & 3.791 & 5.335 & 3 & 0.717 & F3Ia \citep{ardeberg72} \\ 
HD 269661 & 391815407 & 82.70867323 & -69.52483016 & 10.290 & 3.830 & 5.135 & 21 & 4.784 & A0Ia0e: \citep{ardeberg72} \\ 
CD-69   310 & 279957325 & 81.96335132 & -69.01537881 & 10.540 & 3.846 & 5.178 & 13 & 0.101 & F2I \citep{rousseau78} \\ 
HD 269651 & 391813303 & 82.63514108 & -69.15330311 & 10.677 & 3.847 & 5.066 & 2 & 0.098 & F0I \citep{sanduleak70} \\ 
HD  33579 & 31106686 & 76.48129968 & -67.88636971 & 8.947 & 3.854 & 5.716 & 13 & 0.177 & A2Ia$+$ \citep{stock76} \\ 
HD 269781 & 276864037 & 83.59361557 & -67.02321398 & 9.769 & 3.891 & 5.433 & 13 & 0.072 & A0Iae \citep{feast60} \\ 
HD 269604 & 279956577 & 82.13069781 & -68.89881910 & 10.659 & 3.910 & 5.166 & 1 & 0.105 & A1Ia0 \citep{ardeberg72} \\ 
HD 268946 & 31109182 & 76.30091922 & -66.73682529 & 10.242 & 3.946 & 5.398 & 8 & 0.168 & A0Ia \citep{feast60} \\ 
HD 269787 & 276863889 & 83.64317382 & -66.97317331 & 10.705 & 3.955 & 5.241 & 2 & 0.909 & A0Ia0 \citep{feast60} \\ 
HD 269762 & 276869010 & 83.54154627 & -68.98684682 & 10.301 & 3.992 & 5.065 & 4 & 0.176 & A2Ia \citep{stock76} \\ 
SK -69   99 & 179304909 & 79.62570750 & -69.22057018 & 9.776 & 4.022 & 4.973 & 2 & 0.292 & A0I \citep{massey00} \\ 
\enddata
\end{deluxetable*}

\begin{deluxetable*}{lccccccccc}
\tabletypesize{\scriptsize}
\tablecaption{Similar to Table \ref{tab:lmc_fyps} for FYPS identified in the SMC.\label{tab:smc_fyps}}
\tablehead{\colhead{Common Name} & \colhead{TIC Number} & \colhead{R.A.} & \colhead{Dec} & \colhead{$T$} & \colhead{$\log T_{\rm{eff}}$} & \colhead{$\log L/L_\odot$} & \colhead{$N_f$} & \colhead{$f_0$} & \colhead{Lit. Spectral Type}\\
\colhead{} & \colhead{} & \colhead{[deg]} & \colhead{[deg]} & \colhead{[mag]} & \colhead{[K]} & \colhead{$L_\odot$} & \colhead{} & \colhead{d$^{-1}$} & \colhead{}} 
\startdata
 BZ Tuc & 267547804 & 10.43103503 & -73.72330194 & 11.084 & 3.663 & 4.976 & 4 & 0.183 & G0Iab \citep{dorda18} \\ 
{[VA82]} II-2 & 181446366 & 13.97949302 & -72.67515923 & 10.925 & 3.665 & 4.999 & 1 & 0.899 & G1-G6Ia-Iab \citep{gonzalezfernandez15} \\ 
 HV   829 & 181043309 & 12.61998652 & -72.75255907 & 11.210 & 3.771 & 4.968 & 30 & 29.893 & G0Ib \citep{wallerstein84} \\ 
\enddata
\end{deluxetable*}

\begin{deluxetable*}{lccccccccc}
\tabletypesize{\scriptsize}
\tablecaption{Similar to Table \ref{tab:lmc_fyps} for the pulsating stars below $\lum=5.0$ identified in the LMC.\label{tab:lmc_acyg}}
\tablehead{\colhead{Common Name} & \colhead{TIC Number} & \colhead{R.A.} & \colhead{Dec} & \colhead{$T$} & \colhead{$\log T_{\rm{eff}}$} & \colhead{$\log L/L_\odot$} & \colhead{$N_f$} & \colhead{$f_0$} & \colhead{Lit. Spectral Type}\\
\colhead{} & \colhead{} & \colhead{[deg]} & \colhead{[deg]} & \colhead{[mag]} & \colhead{[K]} & \colhead{$L_\odot$} & \colhead{} & \colhead{d$^{-1}$} & \colhead{}} 
\startdata
OGLE BRIGHT-LMC-MISC-498 & 277173427 & 84.20622963 & -69.46036926 & 11.920 & 3.604 & 4.485 & 3 & 1.706 &  \\ 
SP77  45-54 & 425084797 & 82.86744832 & -67.05635779 & 11.070 & 3.612 & 4.890 & 1 & 1.484 &  \\ 
RM 1-586 & 276860287 & 83.25736453 & -66.80148506 & 11.518 & 3.628 & 4.629 & 3 & 0.197 & M: \citep{rebeirot83} \\ 
HD 269680 & 425058250 & 82.84077297 & -70.95708574 & 11.999 & 3.666 & 4.438 & 20 & 0.038 & K5 \citep{cannon36} \\ 
 HV  2447 & 179435530 & 79.87710563 & -68.68603403 & 11.049 & 3.678 & 4.703 & 20 & 0.081 & G1Ia? \citep{feast74} \\ 
2MASS J05405799-6913537 & 404934011 & 85.24170386 & -69.23158000 & 10.058 & 3.712 & 4.790 & 4 & 1.705 &  \\ 
HD 270100 & 389564663 & 86.21073508 & -67.49458871 & 11.032 & 3.717 & 4.783 & 17 & 0.079 & G2:Ia \citep{brunet73} \\ 
 TT Dor & 31183609 & 76.81661563 & -68.88348526 & 11.439 & 3.745 & 4.711 & 15 & 0.096 &  \\ 
HD 269355 & 179309868 & 79.66108562 & -69.76297148 & 10.930 & 3.787 & 4.941 & 1 & 0.857 & F6Ia \citep{rousseau78} \\ 
HD 269860 & 277171889 & 84.17947540 & -69.14824697 & 11.049 & 3.787 & 4.877 & 6 & 1.337 & A3Ia \citep{stock76} \\ 
SK -69  145 & 279957640 & 82.07404881 & -69.07181941 & 11.142 & 3.787 & 4.828 & 7 & 0.684 & F0I \citep{rousseau78} \\ 
TYC 9163-462-1 & 404851968 & 85.08853178 & -69.31158284 & 10.491 & 3.830 & 4.559 & 1 & 0.191 &  \\ 
HD 269790 & 276865044 & 83.66680087 & -67.50265532 & 11.288 & 3.833 & 4.400 & 3 & 0.623 & A5Iab \citep{stock76} \\ 
HD 269809 & 277021897 & 83.92333011 & -69.85413330 & 10.515 & 3.834 & 4.695 & 1 & 0.246 & A4Iab \citep{stock76} \\ 
SK -69  190 & 276869095 & 83.60094300 & -69.00863957 & 11.026 & 3.862 & 4.466 & 4 & 0.201 & A3I \citep{rousseau78} \\ 
SOI 606 & 373679564 & 81.18491967 & -69.63012369 & 10.743 & 3.864 & 4.603 & 6 & 1.727 & A5Ib \citep{stock76} \\ 
{[M2002]} LMC 162129 & 277025094 & 83.80522222 & -69.24292505 & 10.711 & 3.871 & 4.383 & 2 & 0.213 &  \\ 
2MASS J05370943-6919283 & 277172664 & 84.28926870 & -69.32451860 & 10.364 & 3.876 & 4.612 & 5 & 2.038 &  \\ 
SOI 653 & 404851207 & 85.07487544 & -69.46856249 & 11.996 & 3.878 & 4.582 & 6 & 0.978 & A0Ib \citep{stock76} \\ 
CPD-69   505 & 404933915 & 85.50064132 & -69.21248430 & 10.260 & 3.885 & 4.792 & 2 & 1.460 & A1Iab \citep{stock76} \\ 
HD 269316 & 179204591 & 79.38685419 & -69.75209167 & 11.890 & 3.925 & 4.805 & 2 & 0.179 & A1Ia \citep{ardeberg72} \\ 
HD 269638 & 287400952 & 82.50517823 & -67.24358377 & 11.974 & 3.944 & 4.721 & 7 & 0.680 & A2Iab \citep{evans06} \\ 
2MASS J05411804-6929541 & 404965983 & 85.32512871 & -69.49837537 & 11.618 & 3.955 & 4.463 & 1 & 1.408 &  \\ 
HD 268727 & 30275228 & 74.16628121 & -66.74352900 & 11.490 & 3.956 & 4.910 & 2 & 1.648 & A0Ia \citep{sanduleak72} \\ 
TYC 9162-199-1 & 179208856 & 79.49922258 & -69.27074100 & 10.190 & 3.971 & 4.689 & 27 & 0.280 & A2I \citep{massey00} \\ 
SK -69   99 & 179304909 & 79.62570750 & -69.22057018 & 9.776 & 4.022 & 4.973 & 6 & 1.883 & A0I \citep{massey00} \\ 
2MASS J05140600-6920334 & 40796053 & 78.52503557 & -69.34262774 & 10.830 & 4.033 & 4.670 & 2 & 0.260 &  \\ 
\enddata
\end{deluxetable*}

\begin{deluxetable*}{lccccccccc}
\tabletypesize{\scriptsize}
\tablecaption{Similar to Table \ref{tab:lmc_acyg} for pulsators identified in the SMC.\label{tab:smc_acyg}}
\tablehead{\colhead{Common Name} & \colhead{TIC Number} & \colhead{R.A.} & \colhead{Dec} & \colhead{$T$} & \colhead{$\log T_{\rm{eff}}$} & \colhead{$\log L/L_\odot$} & \colhead{$N_f$} & \colhead{$f_0$} & \colhead{Lit. Spectral Type}\\
\colhead{} & \colhead{} & \colhead{[deg]} & \colhead{[deg]} & \colhead{[mag]} & \colhead{[K]} & \colhead{$L_\odot$} & \colhead{} & \colhead{d$^{-1}$} & \colhead{}} 
\startdata
PMMR 103 & 182293477 & 14.96614858 & -72.72644563 & 11.776 & 3.615 & 4.683 & 2 & 0.393 & K/M \citep{prevot83} \\ 
PMMR  58 & 181453919 & 13.44069757 & -72.89416857 & 11.673 & 3.620 & 4.756 & 2 & 0.543 & G5.5-K0Iab \citep{gonzalezfernandez15} \\ 
 BZ Tuc & 267547804 & 10.43103503 & -73.72330194 & 11.084 & 3.663 & 4.976 & 8 & 0.183 & G0Iab \citep{dorda18} \\ 
{[VA82]} II-2 & 181446366 & 13.97949302 & -72.67515923 & 10.925 & 3.665 & 4.999 & 3 & 0.721 & G1-G6Ia-Iab \citep{gonzalezfernandez15} \\ 
OGLE SMC-CEP-921 & 180618091 & 11.72129615 & -72.71437139 & 11.641 & 3.771 & 4.804 & 7 & 0.239 & G0Iab \citep{gonzalezfernandez15} \\ 
 HV   834 & 181446812 & 13.42768977 & -72.28712091 & 11.490 & 3.777 & 4.859 & 1 & 0.447 & F2Ib \citep{wallerstein84} \\ 
TYC 9138-1910-1 & 181446808 & 13.33920655 & -72.28797814 & 11.917 & 3.792 & 4.662 & 2 & 0.273 & F0I \citep{neugent10} \\ 
SK   20 & 180609896 & 12.02630392 & -73.11055357 & 11.118 & 3.841 & 4.594 & 2 & 1.417 & A0-A2Ia \citep{humphreys91} \\ 
SK   58 & 181454398 & 13.51341866 & -72.52908280 & 11.704 & 3.879 & 4.841 & 5 & 0.508 & A3Iab \citep{ardeberg77} \\ 
\enddata
\end{deluxetable*}

\begin{deluxetable*}{lccccccc}
\tabletypesize{\scriptsize}
\tablecaption{Similar to Table \ref{tab:lmc_fyps} for the nonpulsating stars in the LMC sample.\label{tab:lmc_np}}
\tablehead{\colhead{Common Name} & \colhead{TIC Number} & \colhead{R.A.} & \colhead{Dec} & \colhead{$T$} & \colhead{$\log T_{\rm{eff}}$} & \colhead{$\log L/L_\odot$} & \colhead{Lit. Spectral Type}\\
\colhead{} & \colhead{} & \colhead{[deg]} & \colhead{[deg]} & \colhead{[mag]} & \colhead{[K]} & \colhead{$L_\odot$} & \colhead{}} 
\startdata
SP77  39-16 & 231798010 & 78.28535767 & -70.13955726 & 11.651 & 3.618 & 4.624 & K5-M3 \citep{sanduleak77} \\ 
SP77  48-6 & 179576115 & 80.44574373 & -71.32886278 & 9.995 & 3.630 & 5.245 & -- \\ 
2MASS J05344326-6704104 & 276864159 & 83.68030019 & -67.06957314 & 10.502 & 3.630 & 5.038 &  \\ 
SP77  46-10 & 373606593 & 81.13513752 & -68.49853037 & 11.828 & 3.631 & 4.542 & -- \\ 
SP77  37-33 & 40716825 & 78.57863470 & -67.26371224 & 10.887 & 3.632 & 4.901 & K5-M3 \citep{sanduleak77} \\ 
SP77  37-19 & 231743198 & 78.07235270 & -67.29845829 & 10.963 & 3.638 & 4.859 & K5-M3 \citep{sanduleak77} \\ 
HD  33294 & 30855727 & 76.05651200 & -66.43293100 & 11.689 & 3.638 & 4.356 &  \\ 
RM 1-77 & 30322273 & 74.20597669 & -69.80875289 & 10.784 & 3.645 & 4.927 & M: \citep{rebeirot83} \\ 
{[BE74]} 356 & 276793048 & 83.29182691 & -67.73324234 & 11.350 & 3.658 & 4.749 &  \\ 
UCAC2   2515786 & 277101162 & 84.00476338 & -66.84493010 & 11.631 & 3.665 & 4.584 & -- \\ 
 HV  2338 & 31013470 & 76.53689897 & -71.25725540 & 11.989 & 3.677 & 4.348 &  \\ 
 HV   883 & 30526897 & 75.03151958 & -68.45001791 & 11.196 & 3.680 & 4.841 & F8/G0Ia \citep{feast74} \\ 
MACHO  81.9124.8 & 277174595 & 84.10496608 & -69.69209423 & 11.094 & 3.694 & 4.775 & K2III \citep{tisserand13} \\ 
 HV  2369 & 40716446 & 78.47362853 & -67.06345776 & 11.584 & 3.705 & 4.580 & F5?I? \citep{feast74} \\ 
2MASS J05401935-6941226 & 404850155 & 85.08063629 & -69.68962698 & 11.837 & 3.706 & 4.524 &  \\ 
HD 269374 & 179437523 & 80.00908248 & -68.06339429 & 11.074 & 3.711 & 4.761 & M0 \citep{cannon36} \\ 
2MASS J05224569-6950516 & 179583146 & 80.69040104 & -69.84768533 & 11.126 & 3.718 & 4.812 &  \\ 
HD 269879 & 277103906 & 84.19665346 & -66.76277947 & 10.024 & 3.733 & 5.207 & G2Ia \citep{brunet73} \\ 
HD 268828 & 30267198 & 74.16670545 & -69.69898772 & 11.117 & 3.742 & 4.804 & K5 \citep{cannon36} \\ 
HD 268865 & 30527145 & 74.93145620 & -68.52299439 & 10.820 & 3.748 & 4.885 & K0 \citep{cannon36} \\ 
SOI 228 & 391745696 & 82.71533776 & -67.02821180 & 11.996 & 3.767 & 4.435 & F:Ib \citep{stock76} \\ 
CPD-69   496 & 404965397 & 85.24642223 & -69.34317687 & 10.917 & 3.772 & 4.990 & F6Ia \citep{ardeberg72} \\ 
{[BB69]} LMC  4 & 40345470 & 77.28620015 & -68.98539464 & 11.873 & 3.773 & 4.573 &  \\ 
HD 270025 & 389365864 & 85.53888779 & -68.45865498 & 11.365 & 3.781 & 4.730 & F8I: \citep{sanduleak72} \\ 
SOI  49 & 179639601 & 80.57685271 & -66.91640849 & 11.734 & 3.784 & 4.595 & F8Ib \citep{stock76} \\ 
HD 268819 & 30266478 & 73.88524573 & -69.96252378 & 9.589 & 3.784 & 5.381 & F5Ia \citep{macconnell76} \\ 
HD 269868 & 277105522 & 84.14351107 & -67.67537183 & 11.955 & 3.786 & 4.537 & F8I: \citep{sanduleak72} \\ 
HD 269697 & 425083794 & 82.91009524 & -67.46987817 & 9.939 & 3.797 & 5.278 & F6Ia \citep{ardeberg72} \\ 
HD 269662 & 391812839 & 82.71448150 & -69.04960868 & 10.496 & 3.807 & 5.179 & A0Ia \citep{stock76} \\ 
SK -66   94 & 373846742 & 81.60832910 & -66.20319007 & 10.620 & 3.808 & 4.626 & A2I \citep{rousseau78} \\ 
HD 269331 & 179206253 & 79.50763757 & -69.56049032 & 10.114 & 3.810 & 5.307 & A5Ia0 \citep{ardeberg72} \\ 
CPD-69   485 & 404852217 & 85.10567969 & -69.24340124 & 11.095 & 3.811 & 4.854 & A5Ia \citep{rousseau78} \\ 
EM* MWC  112 & 279957197 & 82.09152859 & -68.99674126 & 11.135 & 3.811 & 4.898 & F5Ia \citep{rousseau78} \\ 
HD 269857 & 277108350 & 84.13491608 & -68.90046420 & 10.183 & 3.820 & 5.331 & A9Ia \citep{ardeberg72} \\ 
HD 270050 & 389368279 & 85.62153509 & -67.32832271 & 10.466 & 3.821 & 5.132 & F6Ia \citep{ardeberg72} \\ 
HD 269661 & 391815407 & 82.70867323 & -69.52483016 & 10.290 & 3.830 & 5.135 & A0Ia0e: \citep{ardeberg72} \\ 
SOI 641 & 276934352 & 83.54628200 & -69.42907688 & 11.596 & 3.832 & 4.697 & A9Iab \citep{stock76} \\ 
SK -66  161 & 277101323 & 83.87964235 & -66.75670932 & 11.598 & 3.838 & 4.692 & F5I: \citep{sanduleak70} \\ 
HD 269719 & 276667896 & 83.15242057 & -70.80683528 & 11.773 & 3.842 & 4.650 & F5I \citep{rousseau78} \\ 
SK -68   71 & 179445102 & 80.26515397 & -68.04835179 & 11.193 & 3.848 & 4.381 & A0Ia \citep{stock76} \\ 
HD 269187 & 40716826 & 78.51844441 & -67.26403763 & 11.216 & 3.862 & 4.887 & A9Ia \citep{ardeberg72} \\ 
SK -67  197 & 276865125 & 83.49613191 & -67.53771284 & 11.207 & 3.863 & 4.402 & A0Ia \citep{stock76} \\ 
HD 269139 & 40515236 & 77.61417015 & -69.15325143 & 11.438 & 3.869 & 4.890 & A0Ia \citep{brunet73} \\ 
SOI 605 & 373522149 & 80.84786652 & -69.65661814 & 11.277 & 3.869 & 4.398 & A3Ib \citep{stock76} \\ 
HD 269998 & 404934352 & 85.37537202 & -69.32535614 & 11.730 & 3.871 & 4.816 & F5I: \citep{sanduleak70} \\ 
HD 269735 & 276787778 & 83.25457644 & -69.37479911 & 11.557 & 3.884 & 4.807 & A3Iab \citep{stock76} \\ 
HD 269171 & 231743190 & 78.31668899 & -67.29644975 & 11.637 & 3.887 & 4.762 & A2I \citep{rousseau78} \\ 
SK -67   89 & 179639033 & 80.69628329 & -67.20630491 & 11.817 & 3.896 & 4.677 & A3Ib \citep{stock76} \\ 
HD 269807 & 277100703 & 83.75102371 & -67.02019094 & 10.696 & 3.912 & 5.167 & A5Ia \citep{ardeberg72} \\ 
HD 269841 & 277099961 & 83.96163270 & -67.44151445 & 11.861 & 3.930 & 4.802 & A0Ia \citep{brunet73} \\ 
HD 269678 & 391746270 & 82.80343074 & -67.25221198 & 11.608 & 3.950 & 4.875 & A1I \citep{rousseau78} \\ 
HD 269896 & 277300709 & 84.45473355 & -68.91712291 & 9.264 & 3.953 & 4.938 & ON9.7Ia$+$ \citep{walborn77} \\ 
HD 268949 & 30848631 & 75.80674868 & -68.55982863 & 11.787 & 3.968 & 4.858 & A0Ia \citep{ardeberg72} \\ 
HD 269762 & 276869010 & 83.54154627 & -68.98684682 & 10.301 & 3.992 & 5.065 & A2Ia \citep{stock76} \\ 
HD 268971 & 30759304 & 75.43830598 & -70.59782370 & 11.606 & 4.001 & 4.966 & A0Ia \citep{stock76} \\ 
\enddata
\end{deluxetable*}

\begin{deluxetable*}{lccccccc}
\tabletypesize{\scriptsize}
\tablecaption{Similar to Table \ref{tab:lmc_fyps} for the nonpulsating stars in the SMC sample.\label{tab:smc_np}}
\tablehead{\colhead{Common Name} & \colhead{TIC Number} & \colhead{R.A.} & \colhead{Dec} & \colhead{$T$} & \colhead{$\log T_{\rm{eff}}$} & \colhead{$\log L/L_\odot$} & \colhead{Lit. Spectral Type}\\
\colhead{} & \colhead{} & \colhead{[deg]} & \colhead{[deg]} & \colhead{[mag]} & \colhead{[K]} & \colhead{$L_\odot$} & \colhead{}} 
\startdata
PMMR 135 & 182735393 & 15.65549392 & -72.27361464 & 11.124 & 3.603 & 4.967 & G6.5Ia-Iab \citep{gonzalezfernandez15} \\ 
PMMR  67 & 181453967 & 13.61198900 & -72.88329048 & 11.679 & 3.605 & 4.844 & G7.5Iab \citep{gonzalezfernandez15} \\ 
PMMR 116 & 182300254 & 15.22557921 & -72.86019236 & 10.727 & 3.607 & 5.094 & G4-K1Ia-Iab \citep{gonzalezfernandez15} \\ 
 HV  1685 & 181660065 & 14.11013368 & -73.47314138 & 11.559 & 3.632 & 4.694 & K4Ia-Iab \citep{gonzalezfernandez15} \\ 
SkKM  94 & 181051257 & 13.14856686 & -72.85146782 & 11.815 & 3.633 & 4.677 & K/M \citep{sanduleak89} \\ 
{[M2002]} SMC  41799 & 181878924 & 14.48701346 & -73.56215194 & 11.552 & 3.647 & 4.749 & G1Ia-Iab \citep{gonzalezfernandez15} \\ 
ISO-MCMS J005033.7-731524 & 181042381 & 12.64035126 & -73.25692685 & 11.264 & 3.747 & 4.873 &  \\ 
Flo 675 & 183495106 & 18.85219718 & -73.51233289 & 11.513 & 3.771 & 4.788 & F2I \citep{neugent10} \\ 
 HV   829 & 181043309 & 12.61998652 & -72.75255907 & 11.210 & 3.771 & 4.968 & G0Ib \citep{wallerstein84} \\ 
AzV 197 & 181880034 & 14.50948670 & -72.29904037 & 11.445 & 3.801 & 4.852 & F2Ia \citep{humphreys83} \\ 
RMC  26 & 182730122 & 15.67934353 & -72.12388122 & 11.301 & 3.832 & 4.927 & F0Ia \citep{ardeberg77} \\ 
HD   5277 & 181042742 & 13.21348613 & -73.11490084 & 10.845 & 3.901 & 5.223 & A0Ia \citep{dubois77} \\ 
SK   74 & 181879931 & 14.67178736 & -72.43762088 & 11.442 & 3.921 & 5.016 & A0Ia \citep{ardeberg77} \\ 
\enddata
\end{deluxetable*}

\begin{figure*}[p!]
\gridline{\fig{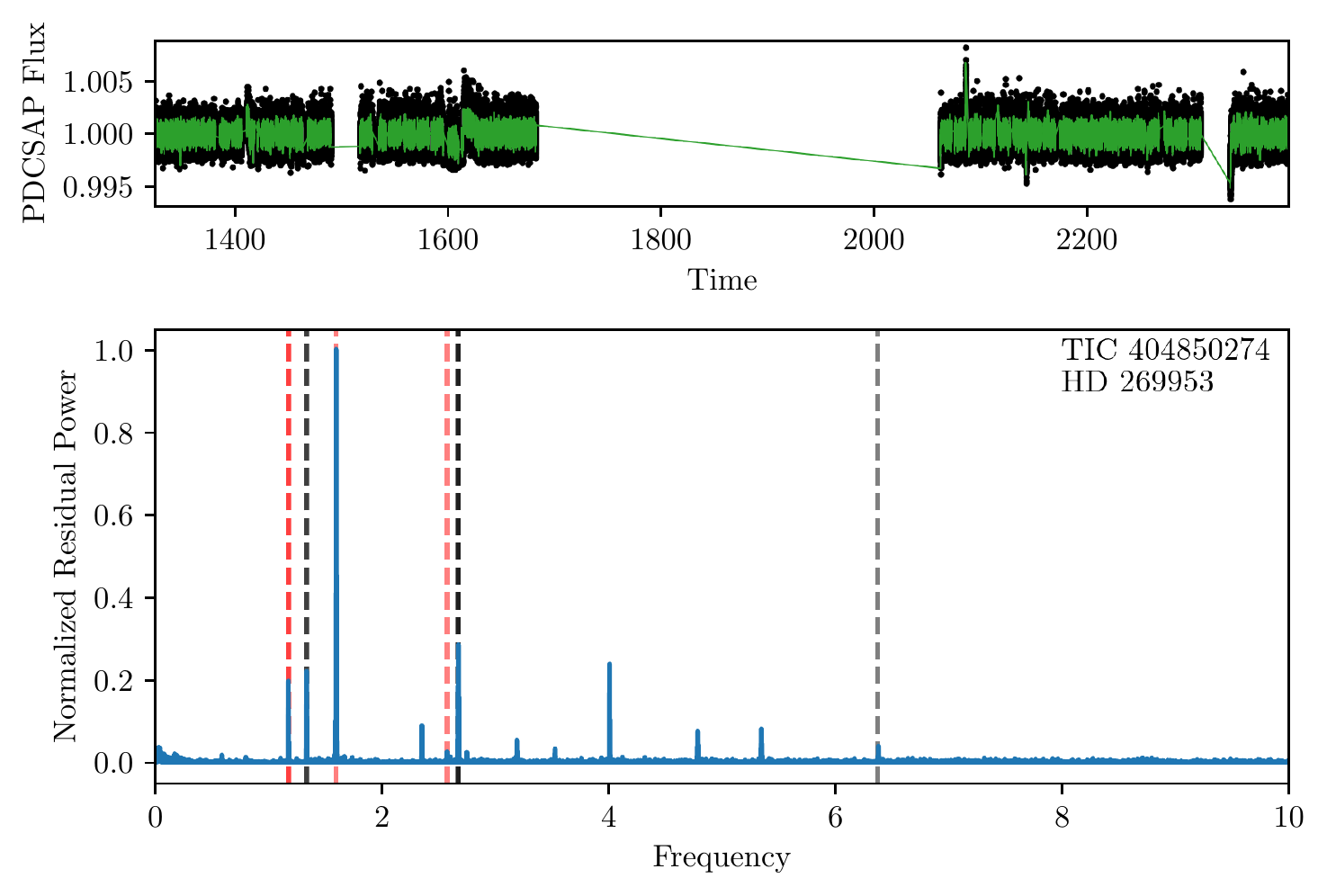}{0.5\textwidth}{}
            \fig{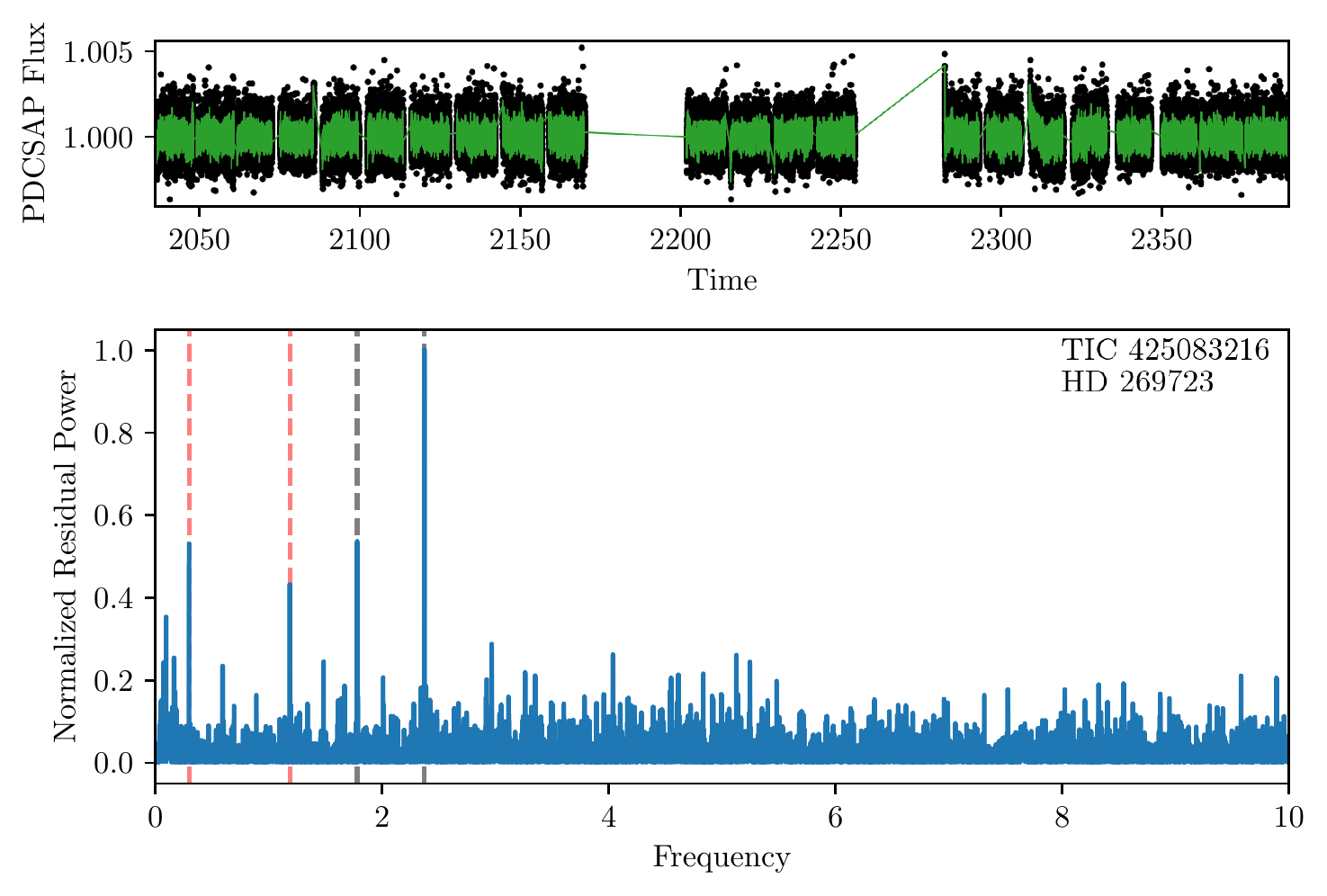}{0.5\textwidth}{}}
            
\gridline{\fig{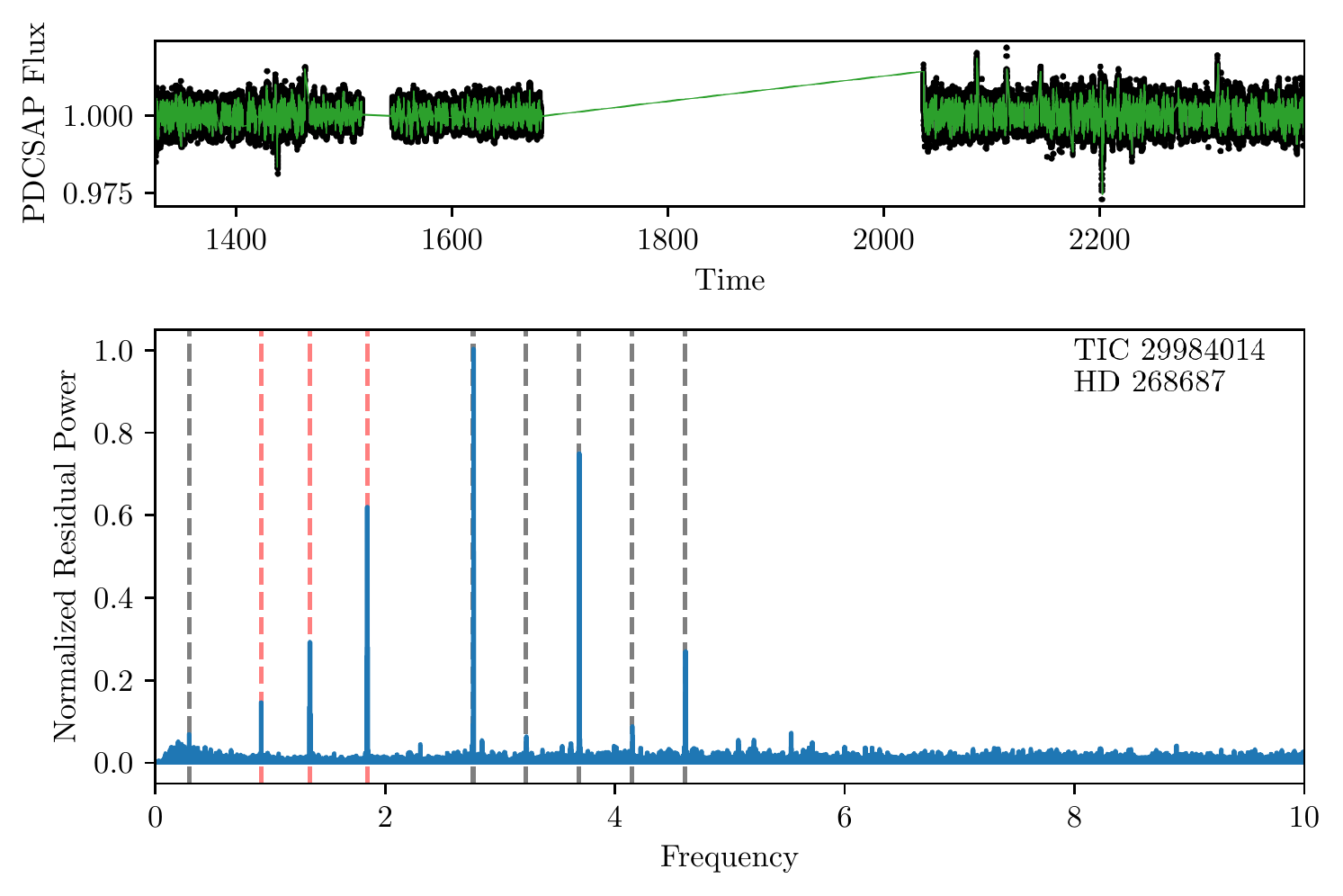}{0.5\textwidth}{}
            \fig{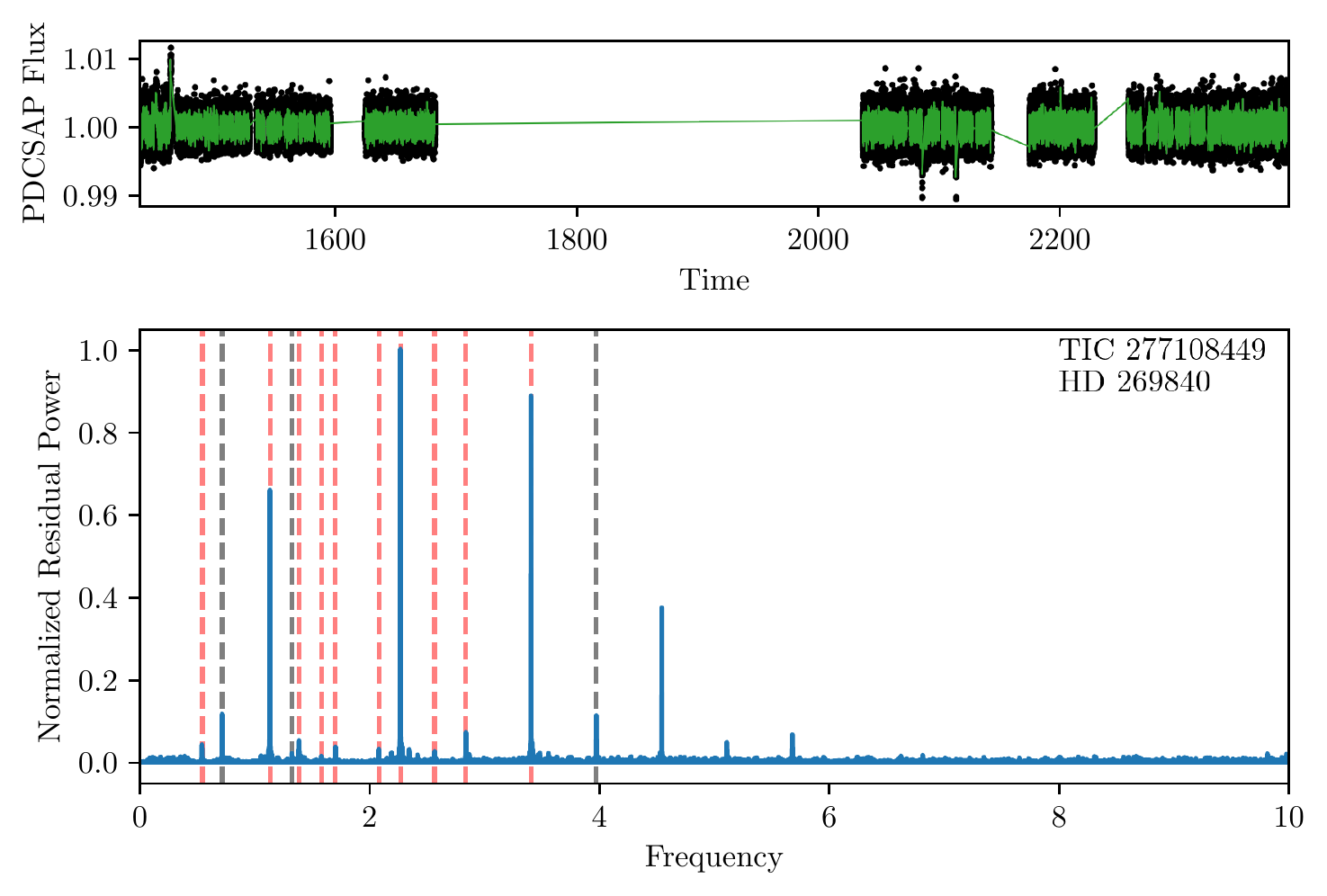}{0.5\textwidth}{}}
            
\gridline{\fig{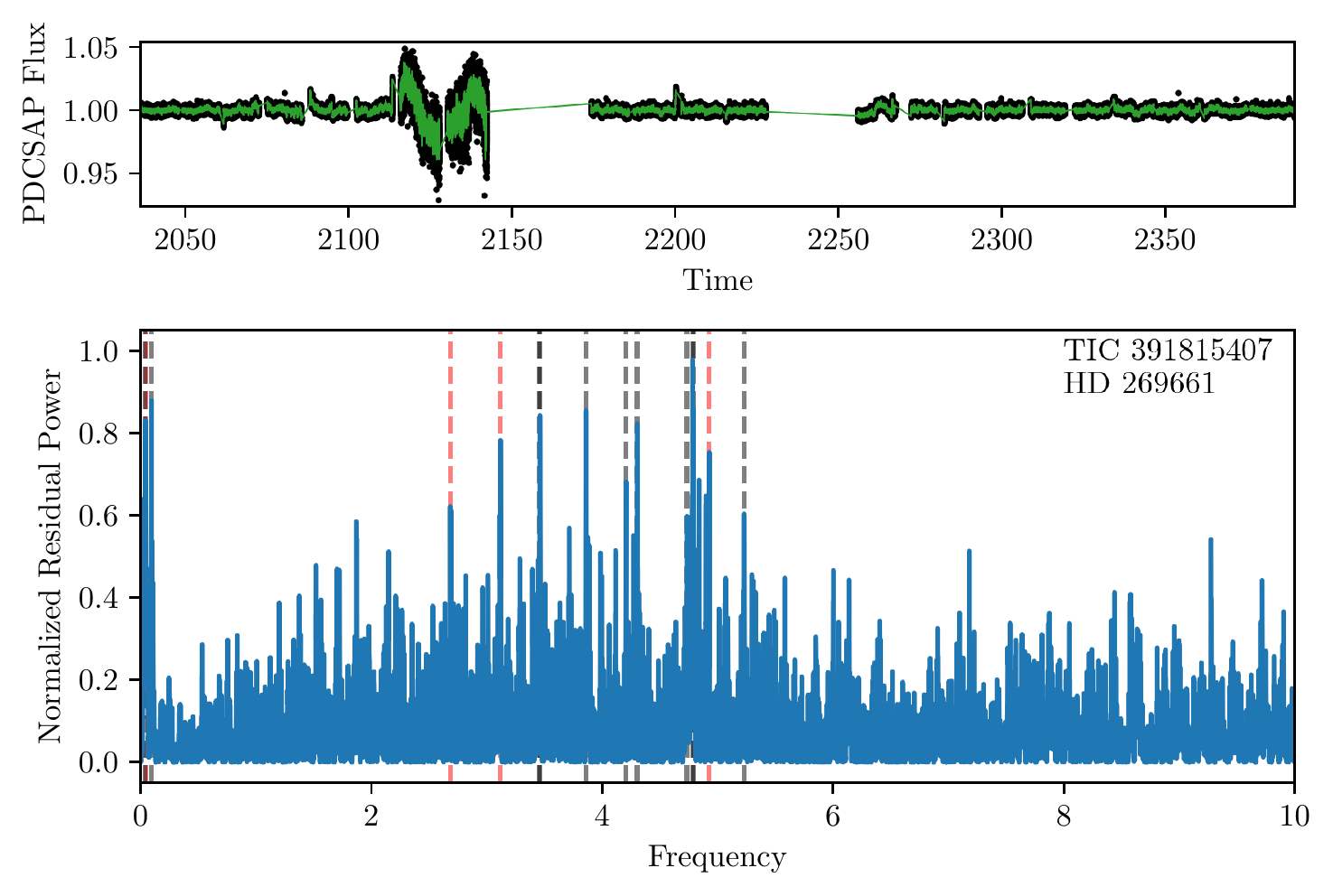}{0.5\textwidth}{}
            \fig{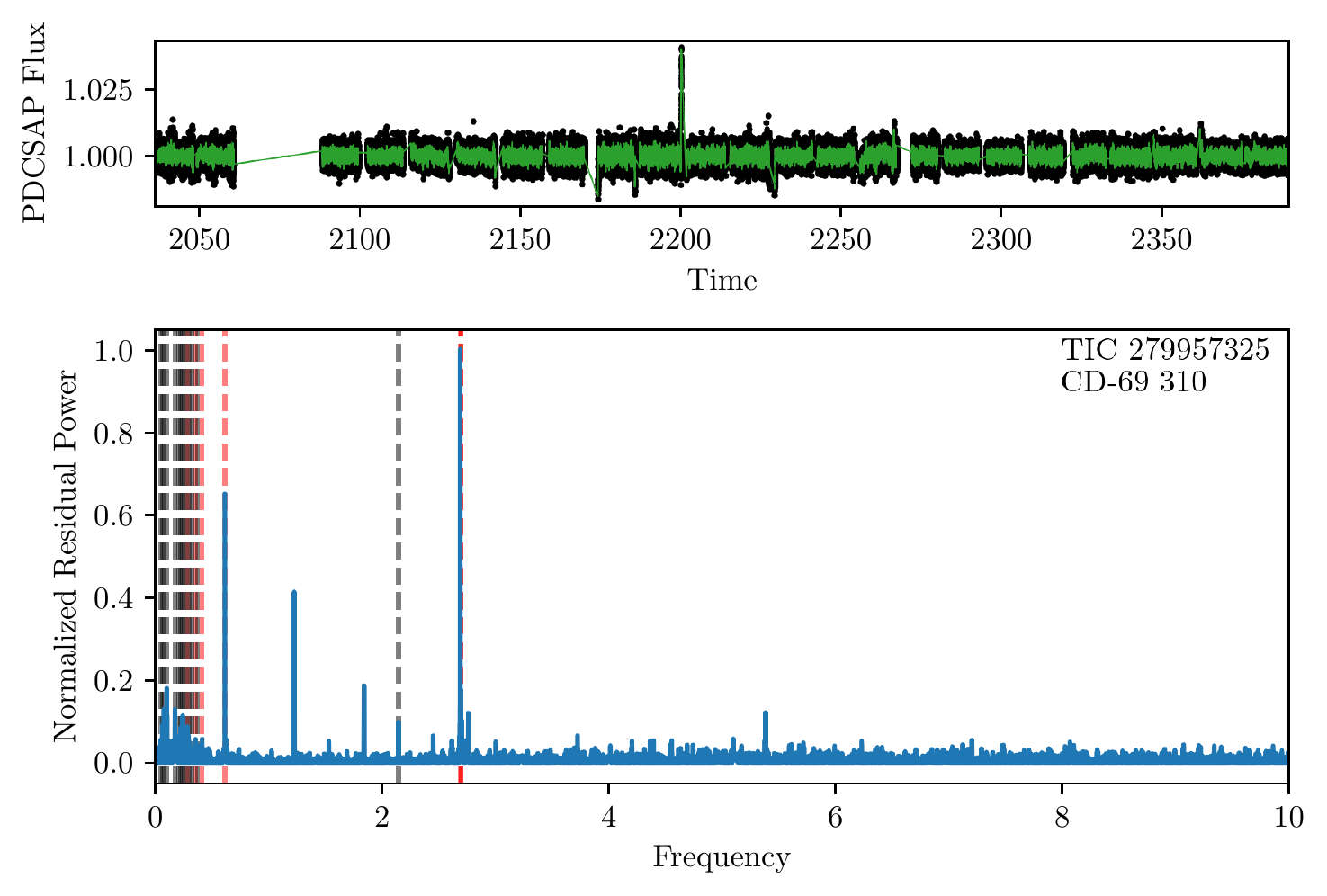}{0.5\textwidth}{}}
            
\caption{Lightcurves and residual power spectra for LMC stars identified as FYPS. Top panels show the normalized \tess~PDCSAP lightcurve as black points, with a 10-minute rolling median in green. Bottom panels show the corresponding Lomb-Scargle periodogram, normalized by Equation \eqref{eq:rednoise} using best-fit parameters derived following \citet{dornwallenstein20b}. Frequencies identified via prewhitening are shown as vertical dashed lines; frequencies associated with nearby contaminants are shown in red, while frequencies associated with the star itself are shown in grey.}\figurenum{1}\label{fig:lmc_fyps}
\end{figure*}
            
\begin{figure*}[p!]
            \gridline{\fig{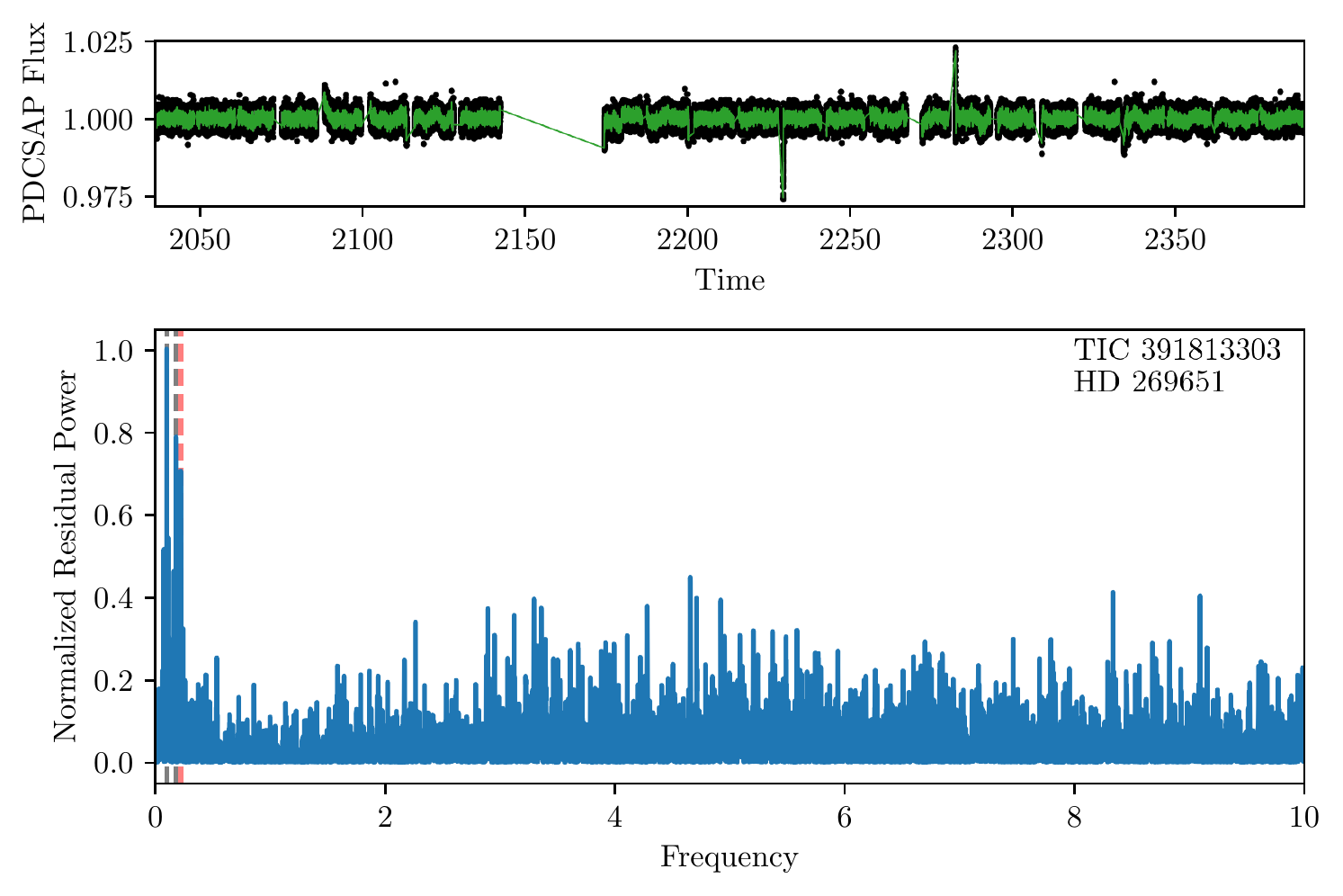}{0.5\textwidth}{}
            \fig{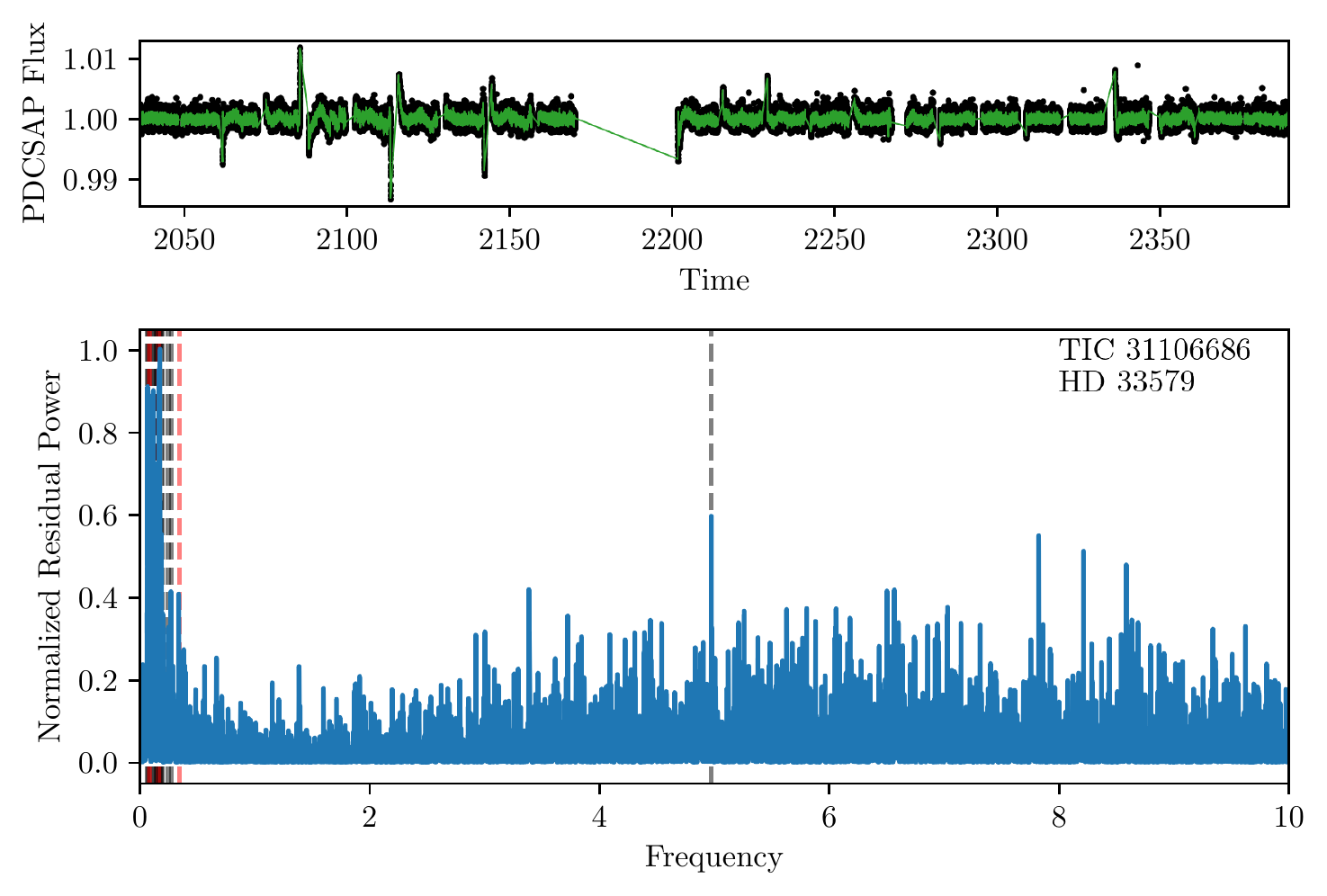}{0.5\textwidth}{}}
            
\gridline{\fig{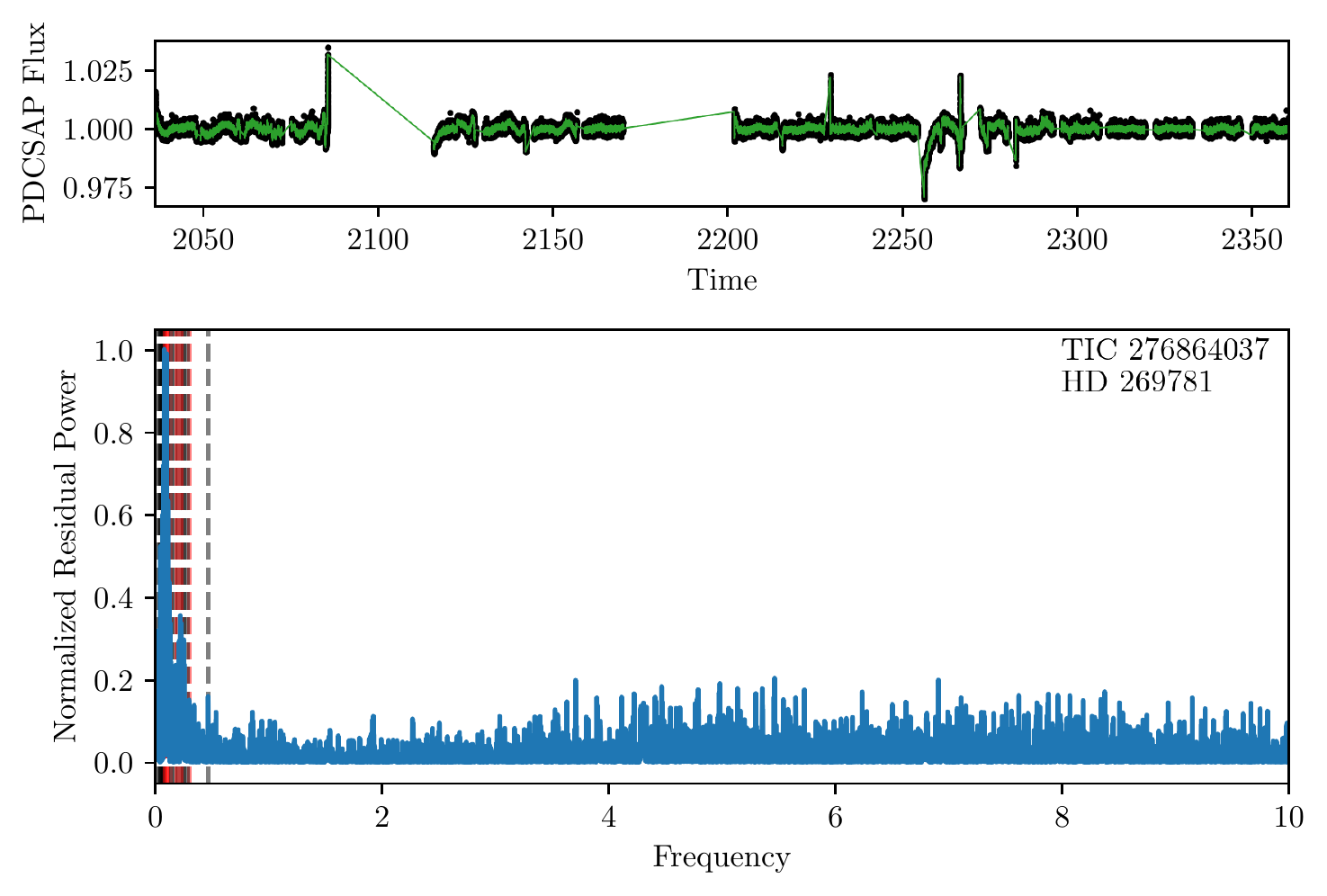}{0.5\textwidth}{}
            \fig{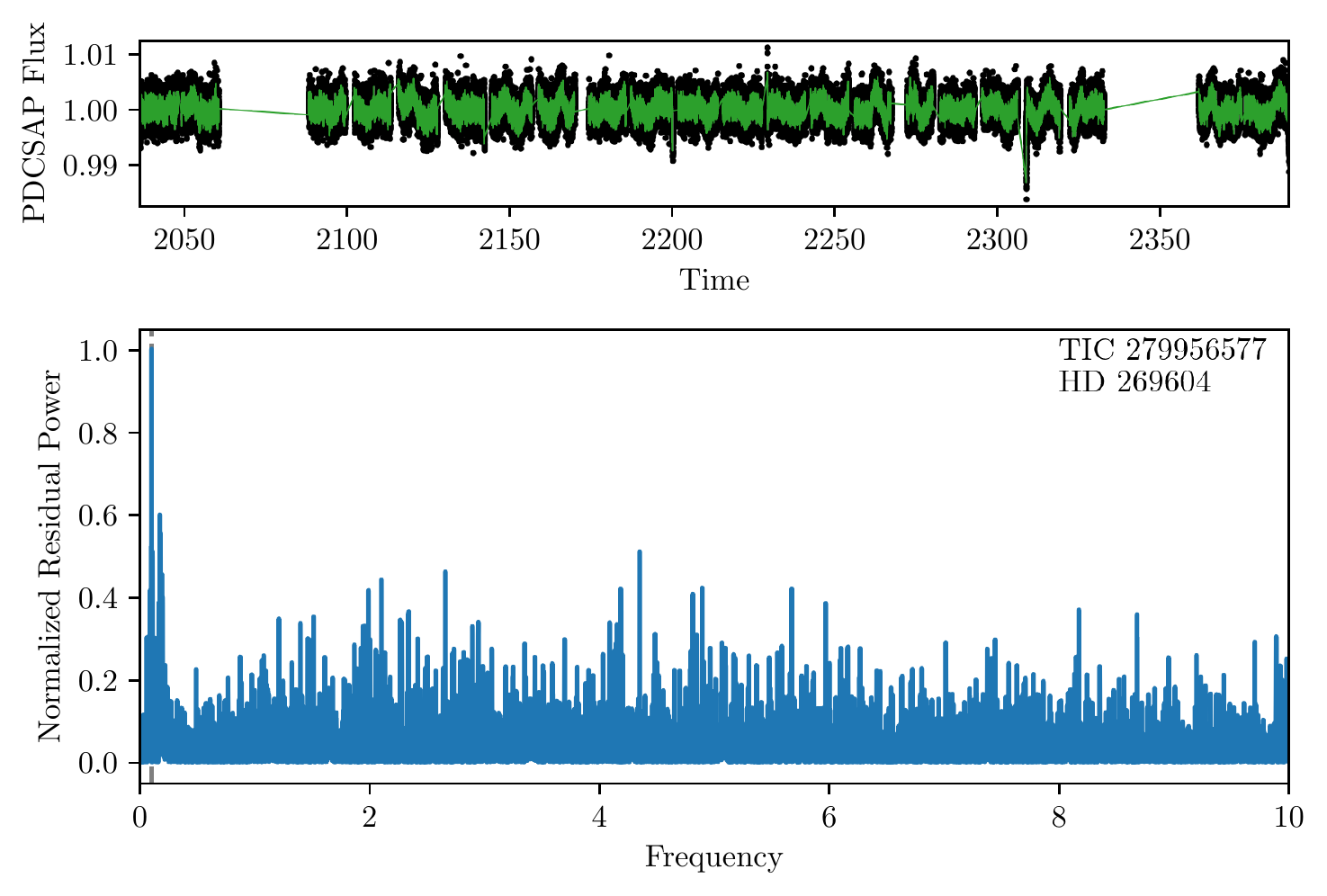}{0.5\textwidth}{}}
            
\gridline{\fig{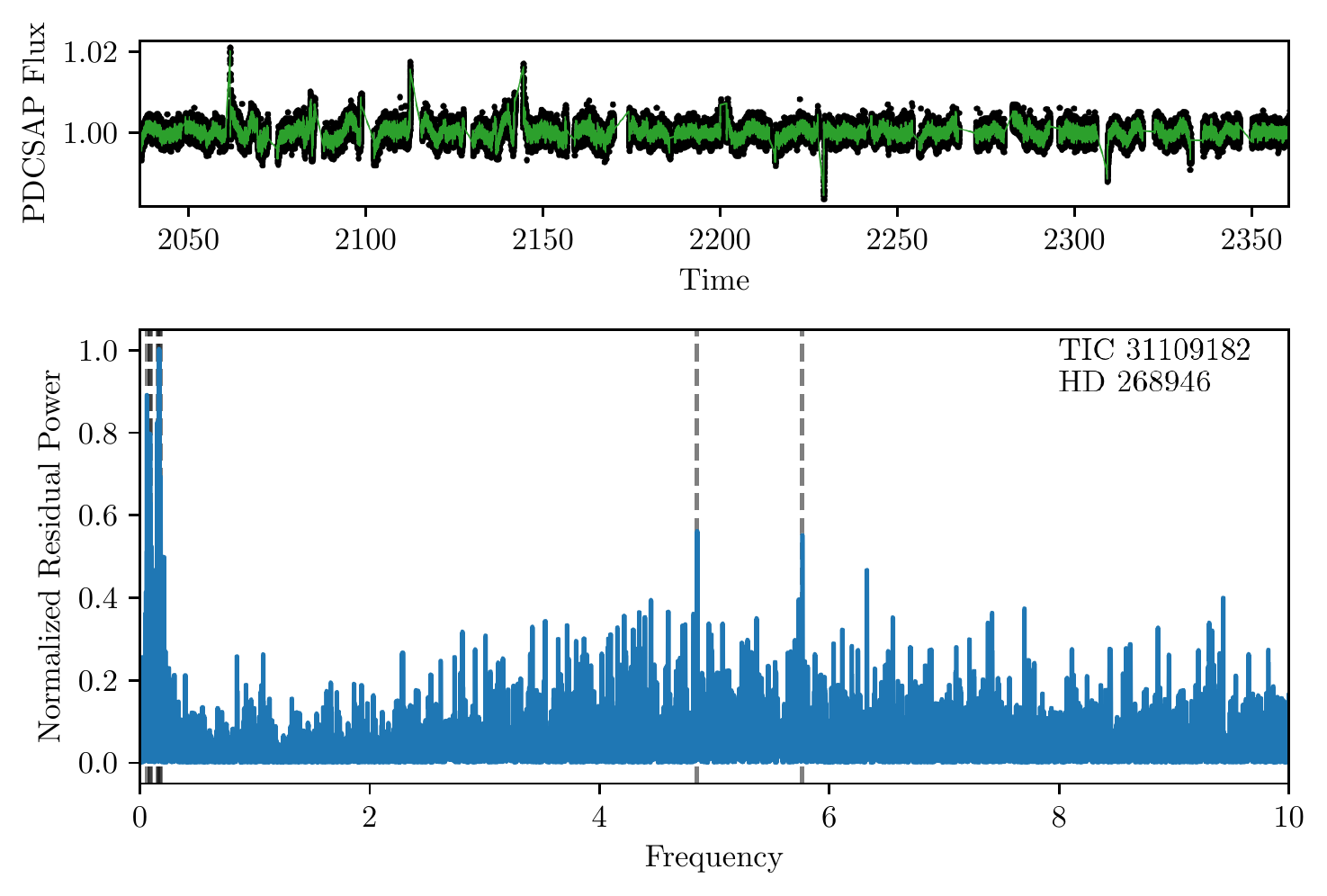}{0.5\textwidth}{}
            \fig{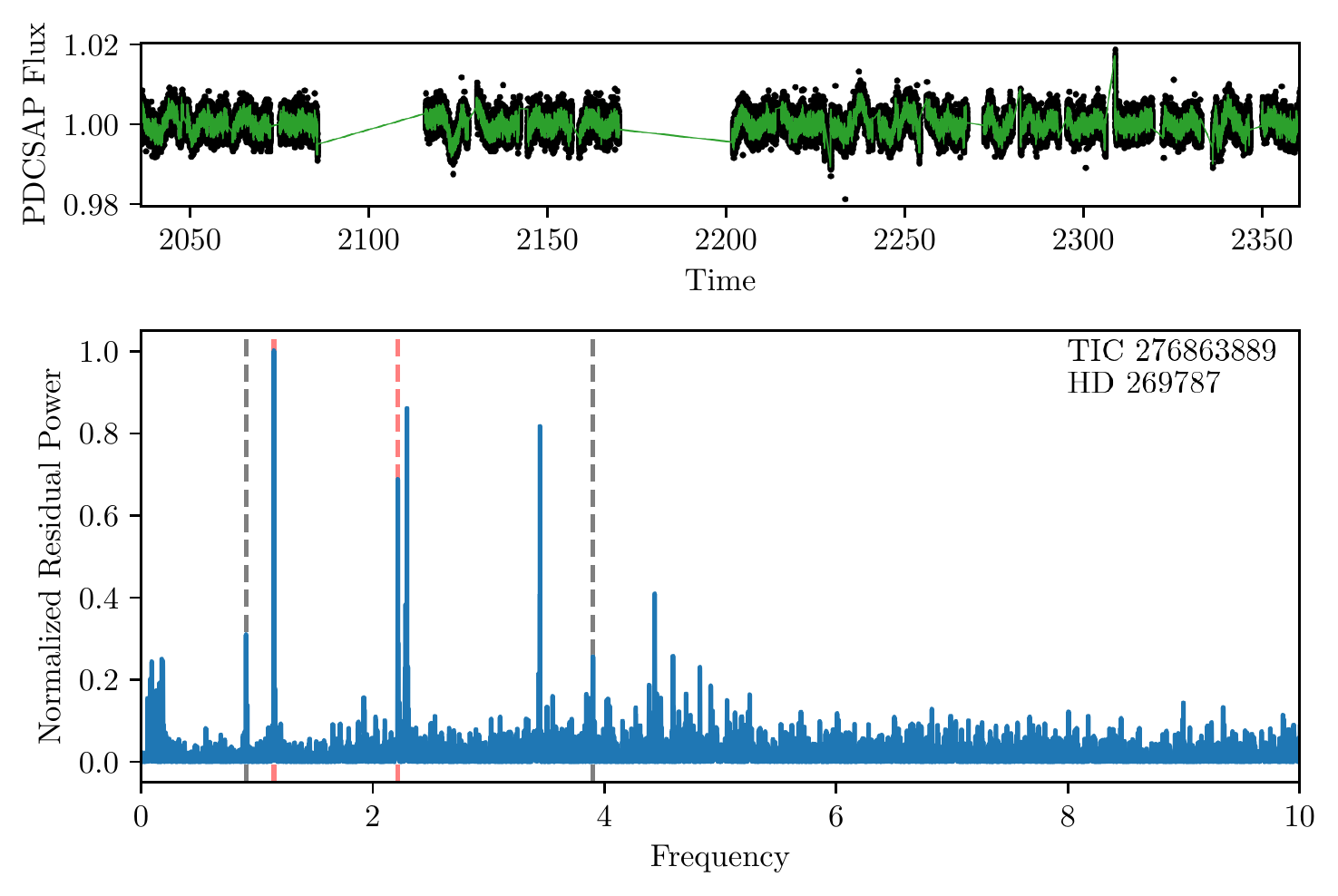}{0.5\textwidth}{}}
            
\caption{{\it cont.}}\figurenum{1}
\end{figure*}
            
\begin{figure*}[p!]
            \gridline{\fig{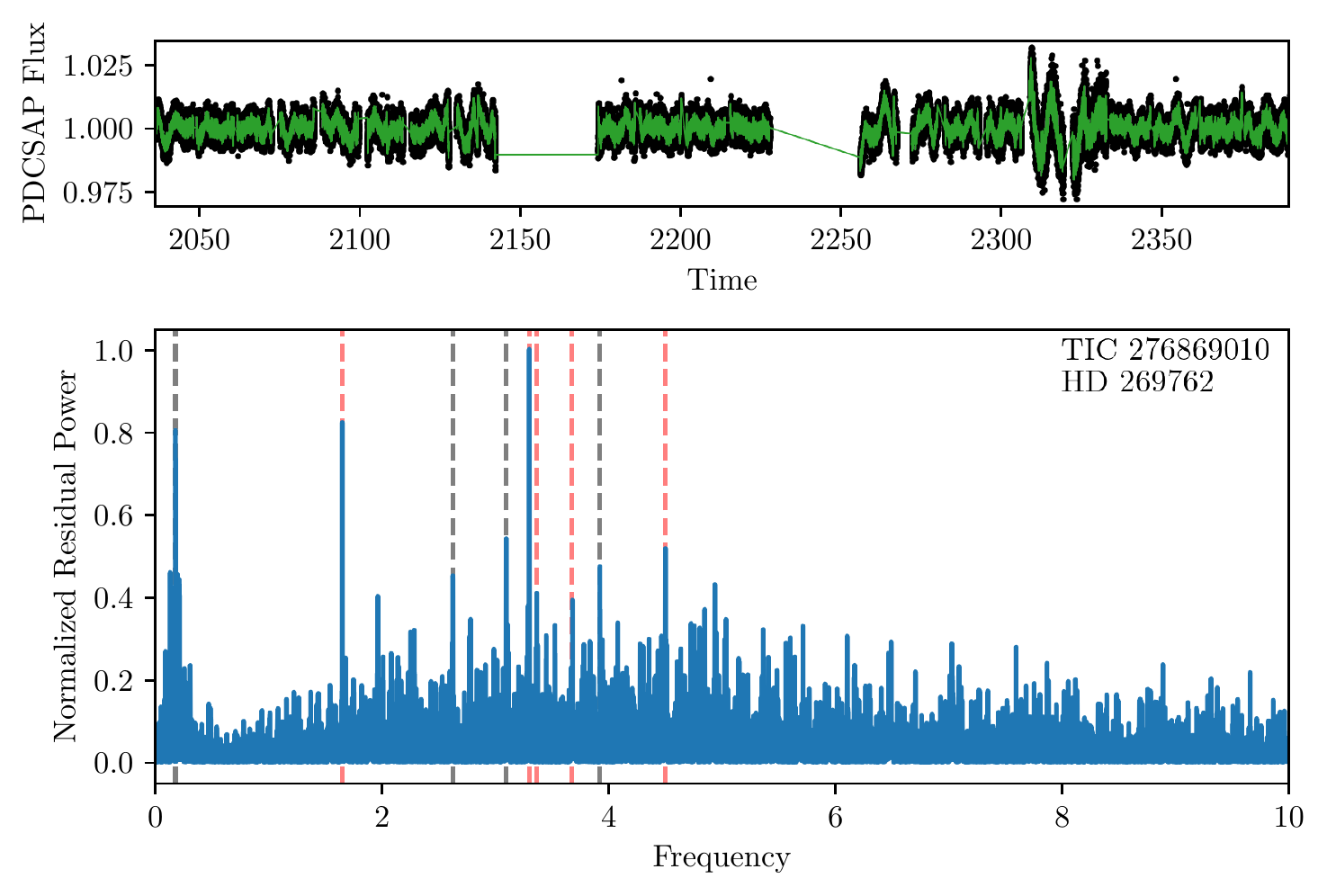}{0.5\textwidth}{}
            \fig{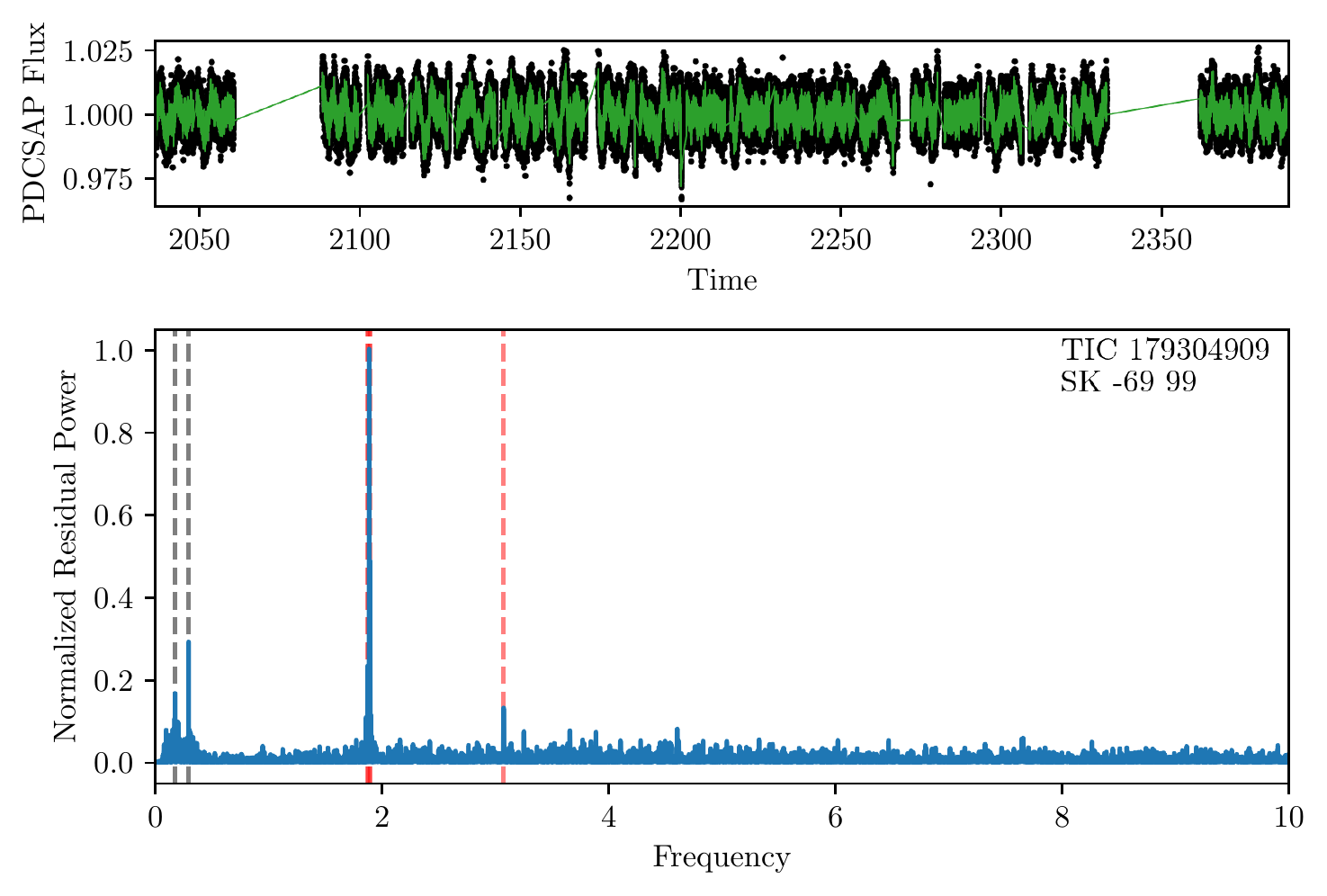}{0.5\textwidth}{}}
            
\caption{{\it cont.}}\figurenum{1}
            \end{figure*}
   
\setcounter{figure}{1}         
\begin{figure*}[p!]
\gridline{\fig{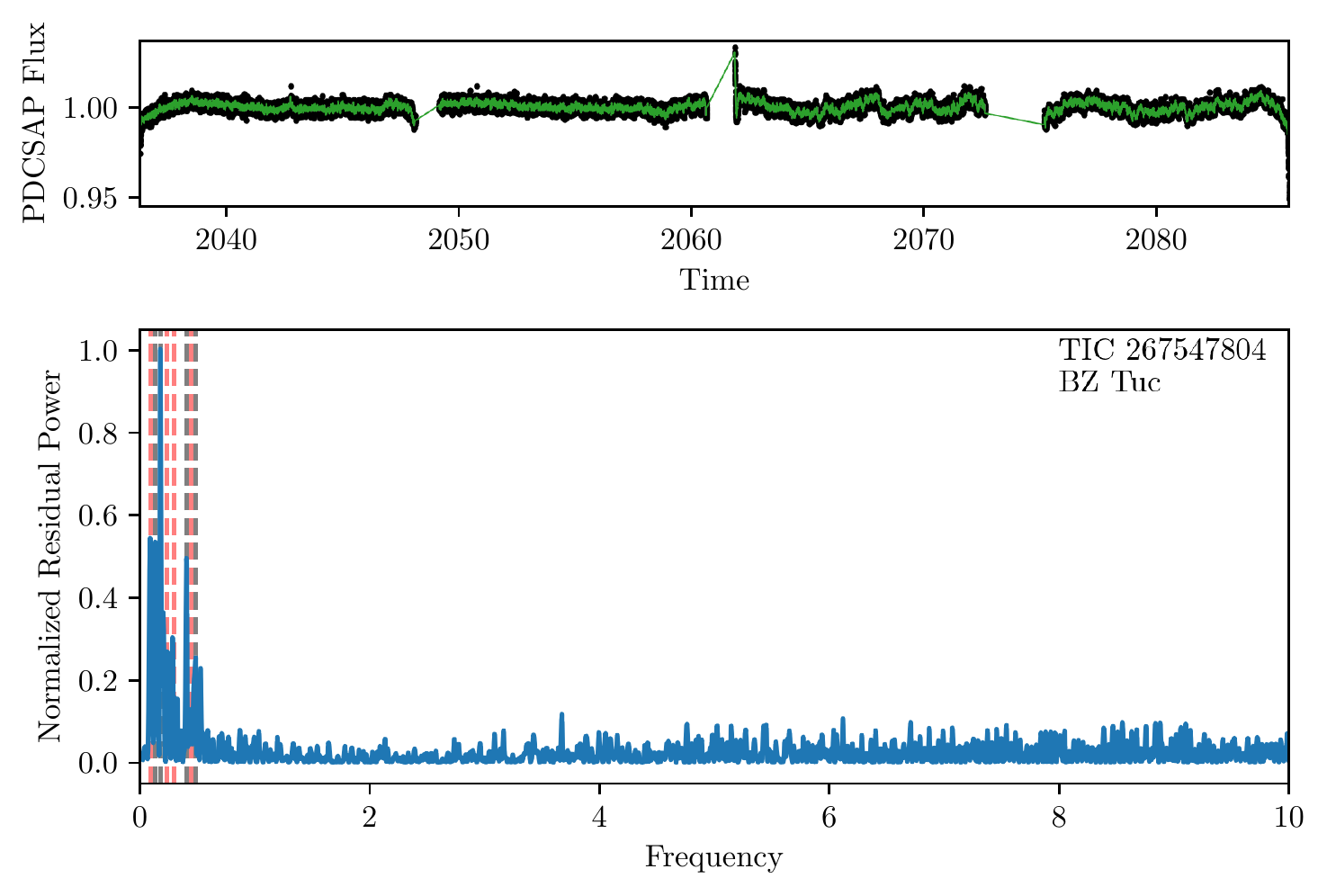}{0.5\textwidth}{}
            \fig{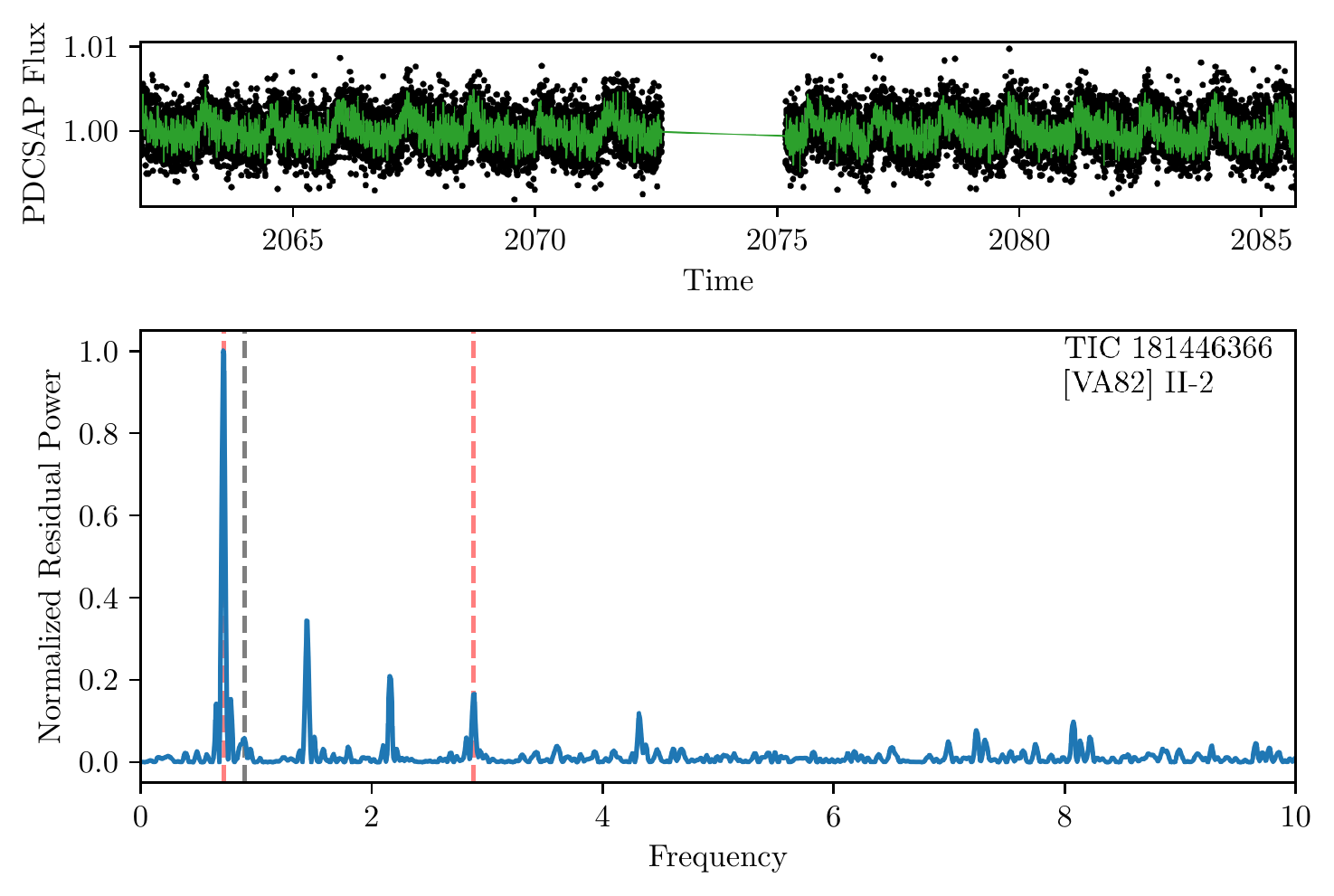}{0.5\textwidth}{}}
            
\gridline{\fig{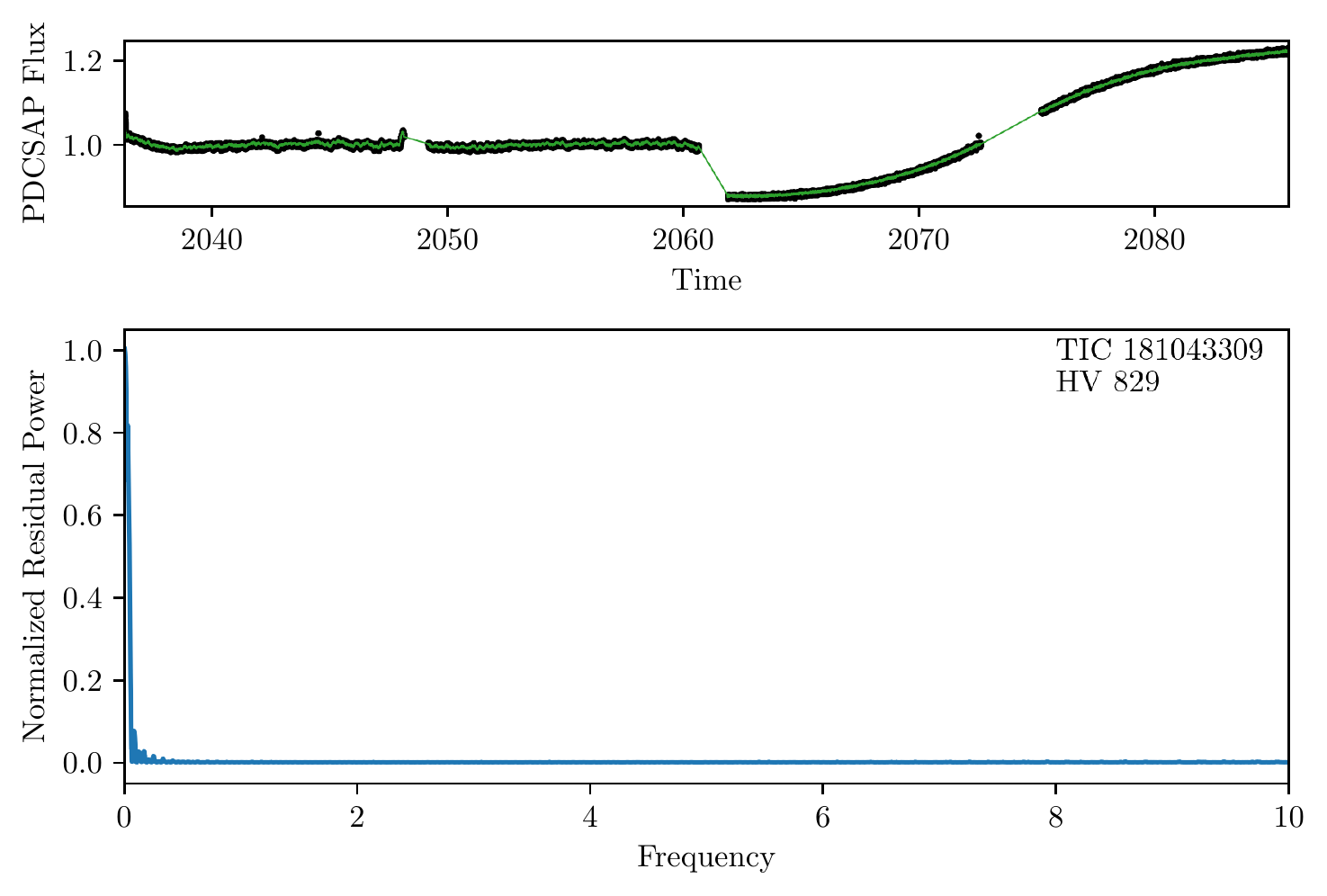}{0.5\textwidth}{}}\caption{Similar to Figure B\ref{fig:lmc_fyps} for FYPS identified in the SMC.}\figurenum{2} \label{fig:smc_fyps}
\end{figure*}

\end{document}